\documentclass[oneside,twocolumn,pdflatex,sn-mathphys-num]{sn-jnl}
\usepackage{graphicx}%
\usepackage{multirow}%
\usepackage{amsmath,amssymb,amsfonts}%
\usepackage{amsthm}%
\usepackage{mathrsfs}%
\usepackage[title]{appendix}%
\usepackage{xcolor}%
\usepackage{textcomp}%
\usepackage{manyfoot}%
\usepackage{algorithm}%
\usepackage{algorithmicx}%
\usepackage{algpseudocode}%
\usepackage{listings}%
\usepackage{multirow}
\usepackage{comment}

\newif\ifshowrebuttal
\showrebuttalfalse % Change to \showrebuttalfalse to turn off

\newcommand{\rebuttal}[1]{%
  \ifshowrebuttal
    \textcolor{red}{#1}%
  \else
    #1%  % Shows normal black text when off
  \fi
}

\newif\ifshowTC
\showTCfalse
\newcommand{\TC}[1]{%
  \ifshowTC
    \textcolor{blue}{#1}%
  \else
    #1%  % Shows normal black text when off
  \fi
}

\newif\ifAWE
\AWEfalse % Change to \AWEfalse to turn off

\newcommand{\AWE}[1]{%
  \ifAWE
    \textcolor{blue}{#1}%
  \else
    #1%  % Shows normal black text when off
  \fi
}

\definecolor{darkred}{RGB}{139,0,0} % Dark red color
\definecolor{darkgreen}{RGB}{0,100,0} % Dark green
\definecolor{amber}{RGB}{205,140,0}

\theoremstyle{thmstyleone}%

\theoremstyle{thmstyletwo}%

\theoremstyle{thmstylethree}%

\usepackage{geometry}
\usepackage[backend=biber, style=numeric-comp, maxnames=5, sorting=none, giveninits=true, firstinits=true]{biblatex}

\DeclareFieldFormat[article,periodical]{volume}{\textbf{#1}}

\DeclareFieldFormat[article]{titlecase}{\MakeSentenceCase*{#1}}

\DeclareNameAlias{default}{family-given}

\DeclareFieldFormat[article]{journaltitle}{\textit{#1}}

\DeclareFieldFormat{doi}{%
  \mkbibacro{doi}\addcolon\space
  \ifhyperref
    {\href{https://doi.org/#1}{\nolinkurl{#1}}}
    {\nolinkurl{#1}}%
}

\begin{document}

\title[Article Title]{Efficient and Accurate Spatial Mixing of Machine Learned Interatomic Potentials for Materials Science}

\author*[1]{\fnm{Fraser} \sur{Birks}}\email{fraser.birks@warwick.ac.uk}

\author[1,2]{\fnm{Matthew} \sur{Nutter}} \email{}

\author[3]{\fnm{Thomas} D \sur{Swinburne}}\email{}

\author[1]{\fnm{James} R \sur{Kermode}}

\affil*[1]{\orgdiv{Warwick Centre for Predictive Modelling, School of Engineering}, \orgname{University of Warwick}, \orgaddress{\street{Library Road}, \city{Coventry}, \postcode{CV4 7AL}, \country{United Kingdom}}}

\affil[2]{\orgdiv{Department of Physics}, \orgname{University of Warwick}, \orgaddress{\street{Library Road}, \city{Coventry}, \postcode{CV4 7AL}, \country{United Kingdom}}}

\affil[3]{\orgdiv{Aix-Marseille Université}, \orgname{CNRS}, \orgaddress{\street{CINaM UMR 7325, Campus de Luminy}, \city{Marseille}, \postcode{13288}, \country{France}}}

\abstract{Machine-learned interatomic potentials can offer near first-principles accuracy but are computationally expensive, limiting their application to large-scale molecular dynamics simulations. Inspired by quantum mechanics/molecular mechanics methods we present ML-MIX, \rebuttal{a CPU- and GPU-compatible} \texttt{LAMMPS} package to accelerate simulations by spatially mixing interatomic potentials of different complexities allowing deployment of modern MLIPs even under restricted computational budgets. We demonstrate our method for ACE, UF3, \rebuttal{SNAP and MACE} potential architectures and demonstrate how linear `cheap' potentials can be distilled from a given `expensive' potential, allowing close matching in relevant regions of configuration space. The functionality of ML-MIX is demonstrated through tests on point defects in Si, Fe and W–He, in which speedups of up to 11× over $\sim$ 8,000 atoms are demonstrated, without sacrificing accuracy. \rebuttal{The scientific potential of ML-MIX is demonstrated via two case studies in W, measuring \AWE{the mobility of $\it{b}$ = $\frac12\langle111\rangle$ screw dislocations} with ACE/ACE mixing and the implantation of He with MACE/SNAP mixing. The latter returns He reflection coefficients which (for the first time) match experimental observations up to an He incident energy of 80~eV -- demonstrating the benefits of deploying state-of-the-art models on large, realistic systems.}}


\maketitle

\section{Introduction}\label{sec1}
Atomistic simulations face two central challenges: modeling quantum-mechanical atomic energies and forces and capturing 
physical processes at large length- and time-scales.
Addressing these challenges under fixed computational budget leads to cost-accuracy trade-offs which have driven methodological development 
over many decades. The most important (and most successful) approximation for electronic interactions is density functional theory (DFT) -- a mean field approach which increases the size of simulations from one or two small atoms to hundreds or even thousands on modern machines \cite{Burke_2012, Jones_2015}. \\

A further level of approximation is to treat electrons implicitly, predicting atomic energies and forces only from nuclear positions and chemical species. For many years, the most accurate models in this category were interatomic potentials with simple functional forms fit to match experimental quantities of interest \cite{Daw_1984, Lennard-Jones_1931, Stillinger_1985, Tersoff_1988} or DFT derived properties \cite{Ercolessi_1994, Brommer_2007}. Many such `empirical' potentials still see considerable use today \cite{Muser_2023}. In the last three decades a new class of interatomic potentials has emerged, which have complex functional forms and parameters fit to DFT data, with an aim to reproduce large swathes of the potential energy surface \cite{Behler_2007, Bartók_2010, Shapeev_2016, Drautz_2019, Deringer_2021, Batatia_2022}. These machine learned interatomic potentials (MLIPs) have achieved remarkable accuracy over a range of applications across materials science and chemistry \cite{Bartók_2018, Kovács_2023, Qamar_2023, Batatia_2024}. Whilst the evaluation costs of empirical potentials and MLIPs both scale linearly with the number of atoms in the system, the complex functional forms of MLIPs result in much larger scaling pre-factors; MLIPs are typically 2-3 orders of magnitude slower in evaluation time than their simple empirical counterparts \cite{Lysogorskiy_2021, Xie_2023}. \\

This cost has slowed the uptake of MLIPs in fields where simulation domains are large and/or timescales are long, which encompasses a large range of phenomena in materials science. It is useful to split such simulations into two categories, those where the entire domain is chemically complex, with atoms involved in bond-making/breaking processes throughout (e.g, amorphous materials \cite{Qian_2023, Grießer_2023}, liquids \cite{Al-Awad_2023, Tian_2023}, high entropy alloys \cite{Byggmästar_2021, Shiotani_2023}), and those where the domain contains only isolated regions of complexity surrounded by atoms in simple environments (e.g, materials defects \cite{Gleizer_2014}, biochemical simulations \cite{Cui_2021}, catalysis \cite{Bramley_2023}). In the second case, simulation cost can in principle be reduced through spatial decomposition: using cheaper, approximate models away from the chemically `important' regions of high complexity. \\

Spatial decomposition has been extensively explored in the context of quantum mechanical/ molecular mechanics modeling (QM/MM), a method which couples an expensive first-principles method (typically DFT) with a linear scaling potential for the complex (QM) and simple (MM) regions of a simulation domain respectively \cite{Warshel_1976, Chen_Ortner_2016, Chen_Ortner_2017, Swinburne_and_Kermode_2017, Bernstein_2009, Kermode_2008, Molani_2024, Grigorev_2023, Wagih_2022}. In this paper, we detail a method that is conceptually similar to QM/MM, but instead combines MLIPs of different complexities (ML/ML). The key difference is that ML/ML simulations aim to run for millions of timesteps, compared to only hundreds or thousands in QM/MM, with each timestep taking 3-6 orders of magnitude less wall time. ML/ML therefore incurs different software challenges: to minimise overhead, operations that take place each timestep must be closely integrated into the molecular dynamics (MD) software. Whilst some recent studies have explored similar `ML/MM' ideas \cite{Hofstetter_2022, Böselt_2021, Immel_2024, Galvelis_2023}, no generic implementation of ML/ML exists. \\

\begin{figure}
    \centering
    \includegraphics[width=\linewidth]{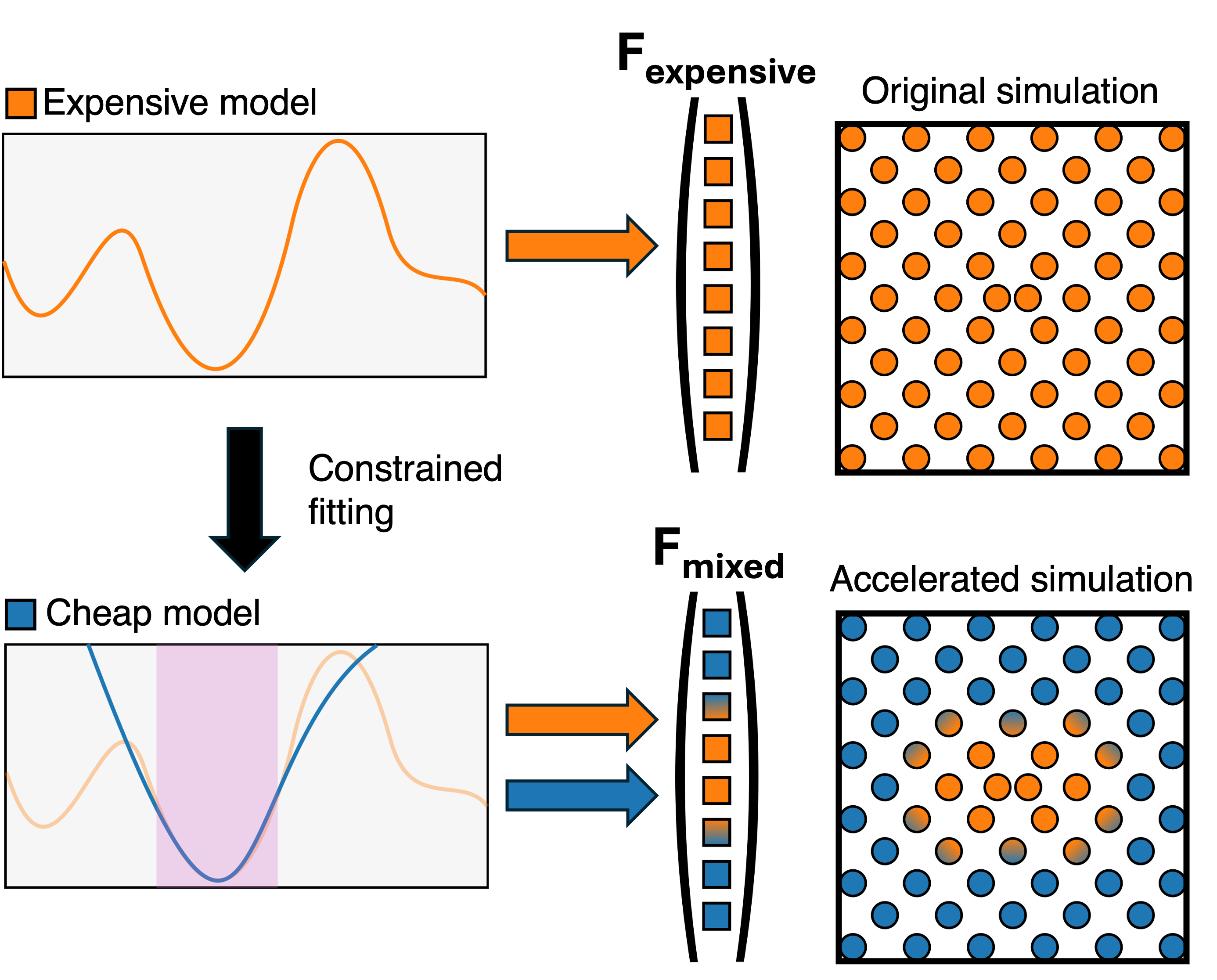}
    \caption{A schematic overview of the simulation acceleration workflow. Left side: the constrained potential fitting process, whereby an accurate expensive potential (orange) is approximated in local regions of potential energy space by a simpler cheap potential (blue). Right side: schematic showing how simulation is accelerated; cheap potential is used to evaluate force components on atoms in `bulk-like' environments.}
    \label{fig:flowchart}
\end{figure}

The central contribution of this work is ML-MIX, a package created to enable the whole workflow of accelerating a simulation within the popular open-source MD software \texttt{LAMMPS} \cite{LAMMPS, type_label_LAMMPS}. Given an `expensive' MLIP, we demonstrate how one can distill a `cheap' linear MLIP that locally approximates the potential energy surface and then show how spatial mixing of the cheap and expensive MLIPs can be used to run accurate simulations at a fraction of the cost. The process is schematically depicted in Fig.~\ref{fig:flowchart}. ML-MIX aims to be as generic as possible; it is able to perform ML/ML and ML/MM simulations with many native \texttt{LAMMPS} potentials out of the box, in parallel, on both CPU \rebuttal{and GPU}. A non-exhaustive list of compatible MLIPs includes \rebuttal{SNAP} \cite{Thompson_2015}, ACE \cite{Drautz_2019}, UF3 \cite{Xie_2023} and \rebuttal{MACE} \cite{Batatia_2022}. An up-to-date full list of tested compatible potentials can be found in the ML-MIX GitHub repository \cite{ML-MIX}. \\

Here we present the results of six ML/ML case-studies, targeting one static and five dynamic quantities of interest in Si, Fe and W, demonstrating that ML-MIX can be used to generate results that agree with those from fully expensive simulations \rebuttal{(and in the final case-study, experiment)} at a fraction of the cost. The first case studies are designed primarily to demonstrate different features of the ML-MIX package. These particularly aim to highlight how ML/ML methods can produce accurate results for simple quantities, even for very small ($\sim 50$ basis function) linear cheap potentials. \rebuttal{We then present two more case studies which were primarily conducted for scientific interest. Both studies concern the application of W as a diverter material in nuclear fusion reactors. In the first of these final two case-studies, we consider the thermally activated glide of screw dislocations in W at different stresses and temperatures. In the second we consider the normally incident deposition of He at different energies into W at 1000 K. In this final case-study, we were able to reproduce (for the first time) experimental observations of the reflection coefficient of He into W for deposition energies up to 80~eV.} \\

\section{Results}\label{sec2}
\subsection{Potential fitting}
% \textbf{Expensive potentials}: 

The starting point for our study are five `expensive' potentials; four linear atomic cluster expansion (ACE) potentials (Si, Fe, W, W -- He) fit using the \texttt{ACEpotentials.jl} package \cite{Witt_2023}, \rebuttal{and one MACE potential created from fine-tuning the mpa-0-medium foundation model to W -- He data using multihead replay fine-tuning \cite{Batatia_2024}.} Table~\ref{tab:expensive_pot_params} shows the fitting parameters of each expensive potential and where to find the corresponding DFT data and parameters. \TC{Additional hyper-parameters for the Fe, W and W -- He ACE potentials are specified in section S1 of the supplementary information}. The root-mean squared per-atom-energy and force errors (RMSEs) for all potentials on their fitting datasets are shown in the upper left quadrant of Table~\ref{tab:all_rmses}. \\

\begin{table}[]
\caption{A summary of expensive potential fitting parameters and source of corresponding DFT data. For the ACE potentials, $\nu$ refers to correlation order and $p$ refers to maximum total polynomial degree. \rebuttal{For the MACE potential, $L$ refers to the message rank.} In all cases, $r_{\mathrm{cut}}$ is the cutoff radius for the local environment.}
\begin{tabular}{l|l|l}
\begin{tabular}[c]{@{}l@{}}Expensive \\ potential\end{tabular} & \begin{tabular}[c]{@{}l@{}}Key fitting \\ parameters\end{tabular} & DFT \\ \hline
Si ACE & \begin{tabular}[c]{@{}l@{}}$\nu = 4$\\ $p = 20$\\ $r_{\mathrm{cut}} = 6.0$~\AA{}\end{tabular} & \cite{Bartók_2018} \\ \hline
Fe ACE & \begin{tabular}[c]{@{}l@{}}$\nu = 3$\\ $p = 20$\\ $r_{\mathrm{cut}} = 5.5$~\AA{}\end{tabular} & \cite{Zhang_2023} \\ \hline
W--He ACE & \begin{tabular}[c]{@{}l@{}}$\nu = 3$\\ $p = 20$\\ $r_{\mathrm{cut}} = 5.0$~\AA{}\end{tabular} & \cite{Nutter_2024, Nutter_2024_supp} \\ \hline
W ACE & \begin{tabular}[c]{@{}l@{}}$\nu = 3$\\ $p = 21$\\ $r_{\mathrm{cut}} = 5.0$~\AA{}\end{tabular} & \begin{tabular}[c]{@{}l@{}}Data: \cite{ml-mix-zenodo}\\ Params: \cite{Nutter_2024, Nutter_2024_supp}\end{tabular} \\ \hline
W--He MACE & \begin{tabular}[c]{@{}l@{}}mpa-0-medium \cite{Batatia_2024}\\ fine tuned\\ 128 L=1 channels\\ 128 L=0 channels\\ 2 layers\\ $r_{\mathrm{cut}} = 6.0$~\AA{}\end{tabular} & \begin{tabular}[c]{@{}l@{}}Pretraining data: \\ MPtraj \cite{mptraj}\\ sAlex \cite{salex_cite_1,salex_cite_2}\\ Fine tuning:\\ Data: \cite{Nutter_2024, Nutter_2024_supp, ml-mix-zenodo}\\ Params: \cite{Nutter_2024, Nutter_2024_supp}\end{tabular}
\end{tabular}
\label{tab:expensive_pot_params}
\end{table}

%\textbf{Cheap potentials}:

The following cheap potentials were generated: (i-ii) Si and Fe linear ACE potentials (fitted with \texttt{ACEpotentials.jl}) with correlation orders of 2 and maximum total polynomial degrees of 10 and cutoffs of 6.0 {\AA} and 5.5 {\AA} respectively, (iii) a W ultra-fast (UF3 \cite{Xie_2023}) potential with a two-body cutoff of 6.0 {\AA} and a 3-body cutoff of 3.5 {\AA} \rebuttal{ and \AWE{(iv) A W linear ACE potential (fitted with \texttt{ACEpotentials.jl}) with a correlation order of 3 and a maximum polynomial degree of 14 and a cutoff of 5.0 {\AA}}}. \rebuttal{We also use (v) the pure W SNAP potential generated by Wood and Thompson~\cite{wood_2017}}. Whilst all these potentials have significantly lower evaluation cost than their expensive counterparts, (iv) and (v) are more sophisticated than (i-iii). This is due to (i-iii) only needing to describe relatively simple bulk interactions, \rebuttal{whilst (iv) and (v) need to also describe surfaces and longer-range strain fields.} In each case, cutoff radii were selected for the cheap potential which were the same or smaller than the expensive potential. \\

% \rebuttal{\textbf{Cheap potentials fit to DFT data} 
The correlation order 3, maximum total polynomial degree 14 W ACE was fit to the same DFT dataset as its expensive counterpart (the correlation order 3 maximum polynomial degree 21 W ACE). RMSE values for this fit are shown on upper right quadrant of Table~\ref{tab:all_rmses}. Whilst the W SNAP potential was also fit to much of the same underlying DFT data \cite{wood_2017}, the full fitting dataset is unpublished and therefore the RMSEs are left out of Table~\ref{tab:all_rmses}. \\

\rebuttal{For $\sim 5000$ atoms, the speedup for the 3, 14 W ACE over the 3, 21 W ACE was $\sim4.4\times$, whilst the speedup of the pure W SNAP potential over the fine-tuned mpa-0-medium model was $\sim29\times$. Note that the MACE/SNAP speedup was measured on GPU, whilst all other quoted speedups were measured on CPU.} \\

%\rebuttal{\textbf{Cheap potentials fit to synthetic data}}
To generate the small cheap potentials (Si and Fe ACE, W UF3), the expensive potentials were used to generate `synthetic' training data focused around a suitable local region of the potential energy surface. In each case, the synthetic fitting data was divided into two types, `soft constraint' data which consisted of bulk-crystal high-temperature MD trajectories, and `hard constraint' data which comprised a set of zero-temperature homogeneous lattice deformations. Constrained linear fitting was used to ensure that elastic constants and lattice parameters closely matched between the cheap and expensive potentials. The constrained fitting scheme is described in detail in section \ref{sec:constrained_fit}.\\

\begin{table*}[]
\centering
\caption{Training RMSE table for both expensive and cheap potentials on the full DFT datasets used for the creation of the expensive potentials and the small synthetic MD datasets used for the creation of the cheap potentials. For the W-He potential, only configurations containing W are used (as the cheap UF3 potential cannot describe He interactions). Note that the synthetic data is generated by the expensive potential, hence the lack of RMSE values in the lower left quadrant. \rebuttal{Energy errors are per-atom. The full DFT data used for the fitting of the the W SNAP is not public, so these RMSEs are not stated.}}
\begin{tabular}{l|l|ll|l|ll}
 &  & \multicolumn{2}{c|}{RMSEs} &  & \multicolumn{2}{c}{RMSEs} \\
 & \begin{tabular}[c]{@{}l@{}}Expensive\\ potentials\end{tabular} & \begin{tabular}[c]{@{}l@{}}Energy \\ (meV)\end{tabular} & \begin{tabular}[c]{@{}l@{}}Force\\ (eV/{\AA})\end{tabular} & \begin{tabular}[c]{@{}l@{}}Cheap\\ potentials\end{tabular} & \begin{tabular}[c]{@{}l@{}}Energy\\ (meV)\end{tabular} & \begin{tabular}[c]{@{}l@{}}Force\\ (eV/{\AA})\end{tabular} \\ \hline
\multirow{5}{*}{Full DFT data} & Si ACE (4, 20) & 2.14 & 0.075 & Si ACE (2, 10) & 106.5 & 0.27 \\
 & Fe ACE (3, 20) & 2.32 & 0.083 & Fe ACE (2, 10) & 18.48 & 0.156 \\
 & W - He ACE (3, 20) & 2.15 & 0.070 & W UF3 & 6549 & 0.16 \\
 & W ACE (3, 21) & 1.90 & 0.066 & W ACE (3, 14) & 4.489 & 0.116 \\
 & W - He MACE & 4.10 & 0.075 & W SNAP & * & * \\ \hline
\multirow{5}{*}{Synthetic MD data} & Si ACE (4, 20) & - & - & Si ACE (2, 10) & 0.21 & 0.045 \\
 & Fe ACE (3, 20) & - & - & Fe ACE (2, 10) & 0.81 & 0.090 \\
 & W - He ACE (3, 21) & - & - & W UF3 & 1.02 & 0.053 \\
\end{tabular}
\label{tab:all_rmses}
\end{table*}

The training energy and force root mean squared error (RMSE) values for the synthetically trained potentials are displayed in the lower right quadrant of Table.~\ref{tab:all_rmses}. \TC{For comparison, potentials with identical hyper-parameters were also fit directly to the same DFT data as the expensive reference. The training RMSE for these fits are shown in the upper right quadrant of Table.~\ref{tab:all_rmses}}. Compared to the corresponding expensive potentials, the speedup for force evaluations with the Si ACE potential was $\sim30\times$, the W UF3 potential was $\sim75\times$, and the Fe ACE potential was $\sim8\times$. Measured speedups were the same for systems of both 8000 and 250,000 atoms. \\

Table~\ref{tab:elastic_constants} shows the elastic constants predicted by the expensive and cheap potentials. For the potentials fit to synthetic data, it also includes rows demonstrating the impact of applying the hard constraint on the elastic constants and lattice parameters. The elastic constants were computed using the \texttt{matscipy} package \cite{matscipy}. In the case of the Si and Fe ACE potentials, adding the hard constraint data led to near-perfect matching of the reference elastic constants. \rebuttal{For the pure W UF3 potential, the $\mathrm{C}_{11}$ and $\mathrm{C}_{44}$ elastic constants got closer to the underlying reference, but $\mathrm{C}_{12}$ was worse, something that can be attributed to the simplicity of the UF3 functional form.} In all cases, the equilibrium lattice parameter of the model became closer to the expensive reference under constrained fitting.

\begin{table*}[] 
\centering
\caption{Elastic constants and lattice parameters according to each potential (expensive, cheap unconstrained, cheap constrained) for Si, Fe and W--He. Elastic constants are fit to stress-strain data using Bayesian linear regression, with the error ($\sigma$) in each value found from the maximum likelihood estimate. For Si and Fe, the addition of the hard constraints to the cheap potential fitting produces elastic constants which match the expensive within $1 \sigma$. For the W--He potential, the constrained fitting improves $\mathrm{C}_{11}$ (matching within $1 \sigma$) and $\mathrm{C}_{44}$ (matching within $2 \sigma$), but leads to a significantly worse $\mathrm{C}_{12}$ than the unconstrained case. In all cases, the constraint dramatically improves the lattice parameter.}
\label{tab:elastic_constants}
\begin{tabular}{cl|llll}
\multicolumn{1}{l}{} &  & $\mathrm{C}_{11}$ (GPa) & $\mathrm{C}_{12}$ (GPa) & $\mathrm{C}_{44}$ (GPa) & a (\AA) \\ \hline
\multicolumn{1}{c|}{Si ACE (4, 20)} & Expensive & 150.8 $\pm$ 1.0 & 56 $\pm$ 2 & 70.5 $\pm$ 0.4 & 5.461 \\
\multicolumn{1}{c|}{Si ACE (2, 10)} & Cheap (unconstrained) & 139 $\pm$ 3 & 81.29 $\pm$ 0.12 & 52.2 $\pm$ 1.9 & 5.661 \\
\multicolumn{1}{c|}{Si ACE (2, 10)} & Cheap (constrained) & 148 $\pm$ 2 & 55 $\pm$ 3 & 69.8 $\pm$ 0.3 & 5.461 \\ \hline
\multicolumn{1}{c|}{Fe ACE (3, 20)} & Expensive & 283 $\pm$ 12 & 154 $\pm$ 6 & 105 $\pm$ 6 & 2.834 \\
\multicolumn{1}{c|}{Fe ACE (2, 10)} & Cheap (unconstrained) & 426 $\pm$ 16 & 364 $\pm$ 2 & 239 $\pm$ 2 & 3.360 \\
\multicolumn{1}{c|}{Fe ACE (2, 10)} & Cheap (constrained) & 284 $\pm$ 11 & 154 $\pm$ 7 & 106 $\pm$ 3 & 2.834 \\ \hline
\multicolumn{1}{c|}{W--He ACE (3, 20)} & Expensive & 520 $\pm$ 14 & 199 $\pm$ 5 & 143 $\pm$ 5 & 3.181 \\
\multicolumn{1}{c|}{W UF3} & Cheap (unconstrained) & 604 $\pm$ 12 & 209 $\pm$ 12 & 180 $\pm$ 6 & 3.136 \\
\multicolumn{1}{c|}{W UF3} & Cheap (constrained) & 527 $\pm$ 11 & 157 $\pm$ 4 & 157 $\pm$ 3 & 3.180 \\ \hline
\multicolumn{1}{c|}{W ACE (3, 21)} & Expensive & 520 $\pm$ 14 & 199 $\pm$ 5 & 143 $\pm$ 5 & 3.180 \\
\multicolumn{1}{c|}{W ACE (3, 14)} & Cheap (unconstrained) & 520 $\pm$ 14 & 196 $\pm$ 5 & 144 $\pm$ 5 & 3.180 \\ \hline
\multicolumn{1}{c|}{\begin{tabular}[c]{@{}c@{}}W--He MACE\\ \end{tabular}} & Expensive & 540 $\pm$ 15 & 215 $\pm$ 8 & 147 $\pm$ 4 & 3.179 \\
\multicolumn{1}{c|}{W SNAP} & Cheap (unconstrained) & 518 $\pm$ 14 & 196 $\pm$ 5 & 144 $\pm$ 5 & 3.180
\end{tabular}
\end{table*}

\subsection{Energy barrier for Si vacancy migration}
\begin{figure}
    \centering
    \includegraphics[width=\linewidth]{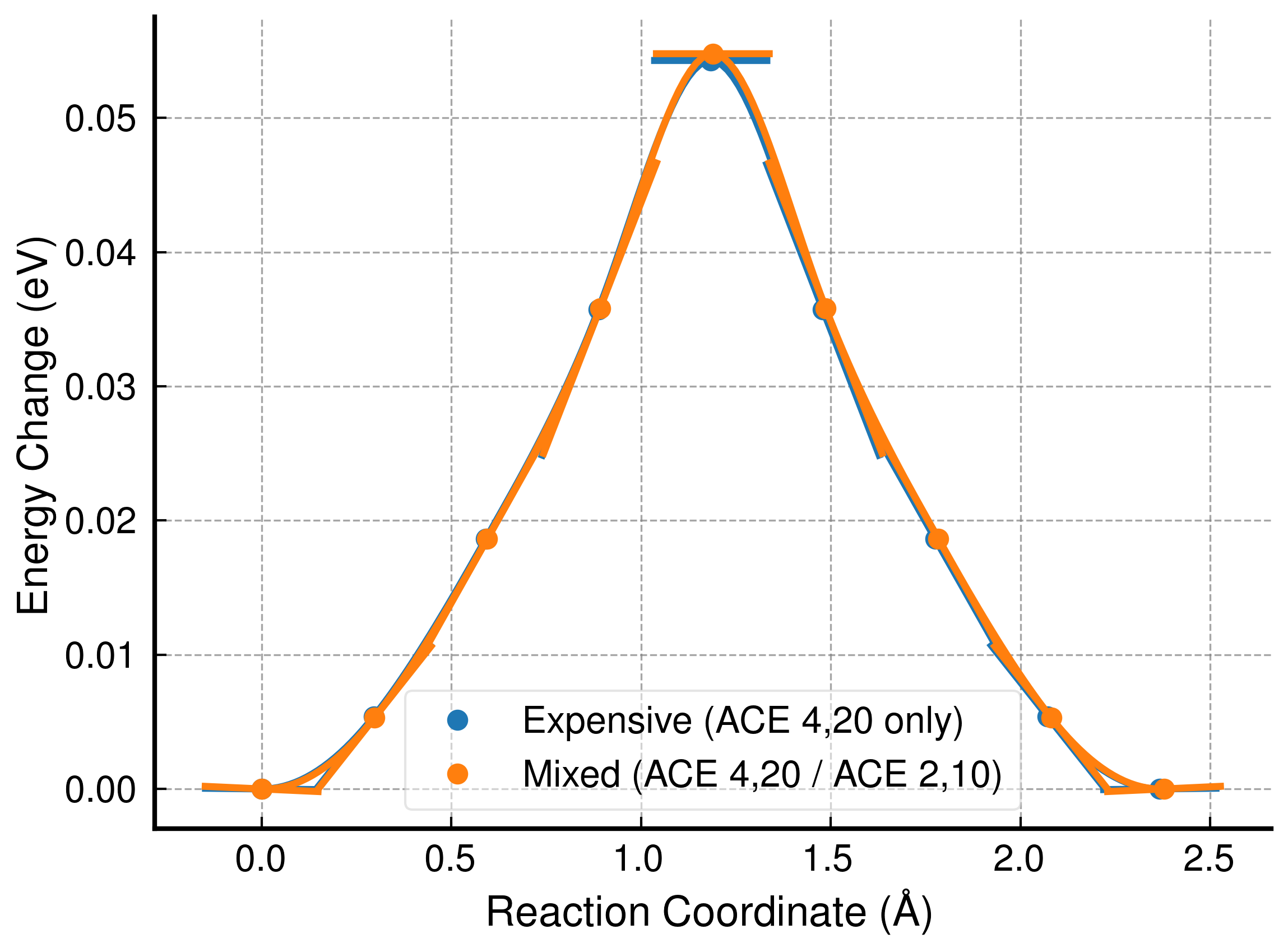}
    \caption{The energy barrier associated with vacancy migration in Si, obtained through all-expensive (blue curve) and ML/ML (orange curve) nudged elastic band simulations. For the ML/ML simulation, the final energies were obtained through one-shot energy evaluations of the relaxed structures with the expensive potential. The all-expensive energy barrier (0.0543 eV) agrees with the ML/ML energy barrier (0.0548 eV) within 1 meV. The tangent lines arise from the projected forces.}
    \label{fig:NEB_plot}
\end{figure}
We first selected a simple zero temperature (static) quantity of interest for a materials defect. We measured the energy barrier for vacancy migration in a $10\times10\times10$ (8000 atom) block of Si for both a fully-expensive simulation and an ML/ML simulation where the expensive potential was localised around the vacancy. For the ML/ML NEB, the combined relaxation was performed using ML-MIX. Fig.~\ref{fig:NEB_plot} shows the results from this investigation. The energy barrier obtained through the ML/ML simulation (0.0548 eV) agrees with the all-expensive reference energy barrier (0.0543 eV) to within 1 meV. The NEB relaxation ran $\sim 5 \times$ faster, close to the theoretical `zero-overhead' value of $5.90 \times$ expected from the fraction of expensive atoms in the ML/ML simulation.

\subsection{Stretched bond in silicon} \label{sec:si_stretched_bond}
\begin{figure}
    \centering
    \includegraphics[width=1.0\linewidth]{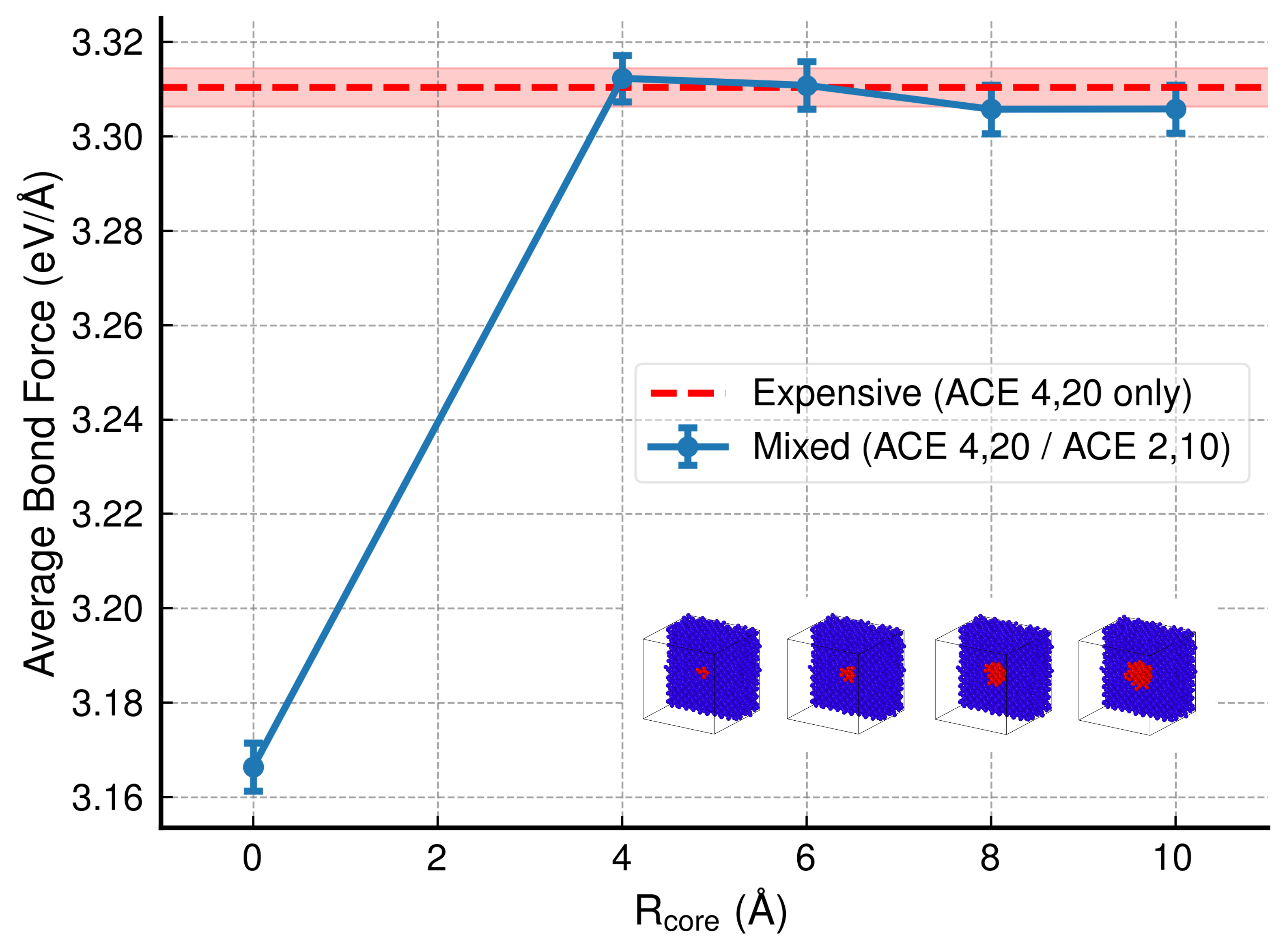}
    \caption{Average force on stretched bond in silicon over 100~ps at 300~K for different expensive potential radii $r_{\mathrm{core}}$ \TC{(blue points)}. \rebuttal{Error bars represent the standard error in bond force measured over this time, with samples taken every 15~fs.} For $r_{\mathrm{core}} = 0$, which corresponds to using the cheap potential everywhere, the average bond force measured does not match the all-expensive $NVE$ reference \TC{(red dashed line)}. Once the expensive potential is introduced at and around the stretched bond, this difference vanishes and agreement between the mixed simulation and the $NVE$ reference is within statistical error.}
    \label{fig:2_10_si}
\end{figure}

Turning to a dynamical example, we investigated the quantity of interest proposed as a test for the quality of different QM/MM techniques by Bernstein et al \cite{Bernstein_2009} --- the average force on a single stretched and rigidly fixed Si bond in a block of bulk Si at 300~K. \\

In a $10\times 10 \times 10$ (8000 atom) supercell of bulk Si, a bond was stretched an additional 0.1~{\AA} along its length and fixed. The average force on this stretched bond recorded over 100~ps was then found, both for all-expensive reference and a number of ML/ML simulations, where the impact of changing the radius of the expensive potential region around the stretched bond atoms $r_\mathrm{core}$ on the measured average force was investigated. The first simulation, at $r_\mathrm{core} = 0$~{\AA}, corresponded to an all-cheap simulation. Further simulations were conducted for $r_\mathrm{core} = 4$~{\AA}, $r_\mathrm{core} = 6$~{\AA}, $r_\mathrm{core} = 8$~{\AA} and $r_\mathrm{core} = 10$~{\AA}.  \rebuttal{A core region size of 2~{\AA} was not tested as a core radius smaller than the Si-Si bond length of 2.35~{\AA} results in no expensive atoms, and thus an identical result to the $r_\mathrm{core} = 0$~{\AA} case.} \\

The results for these simulations are presented in Fig.~\ref{fig:2_10_si}. It can be seen that as the expensive potential replaces the cheap potential in the core ($r_\mathrm{core} > 0$), the forces jump to closely matching the $NVE$ reference. Images of the simulation domain highlighting the atoms treated expensively around the fixed bond are shown on the insets in Fig.~\ref{fig:2_10_si}. \\

The theoretical and measured speedups are shown in Table~\ref{tab:si_speedups}. It can be seen that in serial, the measured speedups match the theoretical, whilst in parallel the values are lower due to imperfect load balancing and parallel overheads. As $r_{\mathrm{core}}$ increases, the number of expensive atoms in the simulation increases, and the overall speedup decreases. Load balancing is further discussed in section S2 of the supplementary information.

\begin{table}[]
\caption{Speedup table for mixed simulations in Si. \TC{The measured serial speedup closely tracks the theoretically predicted value, whilst the parallel speedup is worse than expected for small $r_{core}$ due to imperfect load balancing.}}
\label{tab:si_speedups}
\begin{tabular}{l|lll}
Simulation & \begin{tabular}[c]{@{}l@{}}Theoretical\\ Speedup\end{tabular} & \begin{tabular}[c]{@{}l@{}}Serial\\ Speedup\end{tabular} & \begin{tabular}[c]{@{}l@{}}Parallel\\ Speedup\end{tabular} \\ \hline
Si: $r_{\mathrm{core}}=4$ & 12.93 & 12.89 & 8.89\\
Si: $r_{\mathrm{core}}=6$ & 10.79 & 10.73 & 10.10\\
Si: $r_{\mathrm{core}}=8$ & 8.55 & 8.55 & 8.89\\
Si: $r_{\mathrm{core}}=10$ & 6.72 & 6.72 & 6.88
\end{tabular}
\end{table}

\subsection{Diffusion coefficients}
Next, we moved to a less artificial dynamical quantity of interest. Here, we consider the diffusion of a dumbbell self-interstitial in Fe and an interstitial He impurity in W, both of which have been extensively studied \cite{Wang_2012, Hammond_2018, Willaime_2005, Marinica_2011}. \rebuttal{For each of these quantities, an atomistic domain of ($\sim 8000$ atoms) was used. Whilst domains of this size are not necessary for accurate quantities of interest in these simple cells (we found one can attain converged diffusion coefficients in cells of $\sim 500$ atoms and smaller, depending on the defect), we selected this size primarily for demonstration purposes; it is around the minimum cell size beyond which we believe ML/ML methods are worth implementing.}\\

\begin{figure*}
    \centering
    \includegraphics[width=0.95\linewidth]{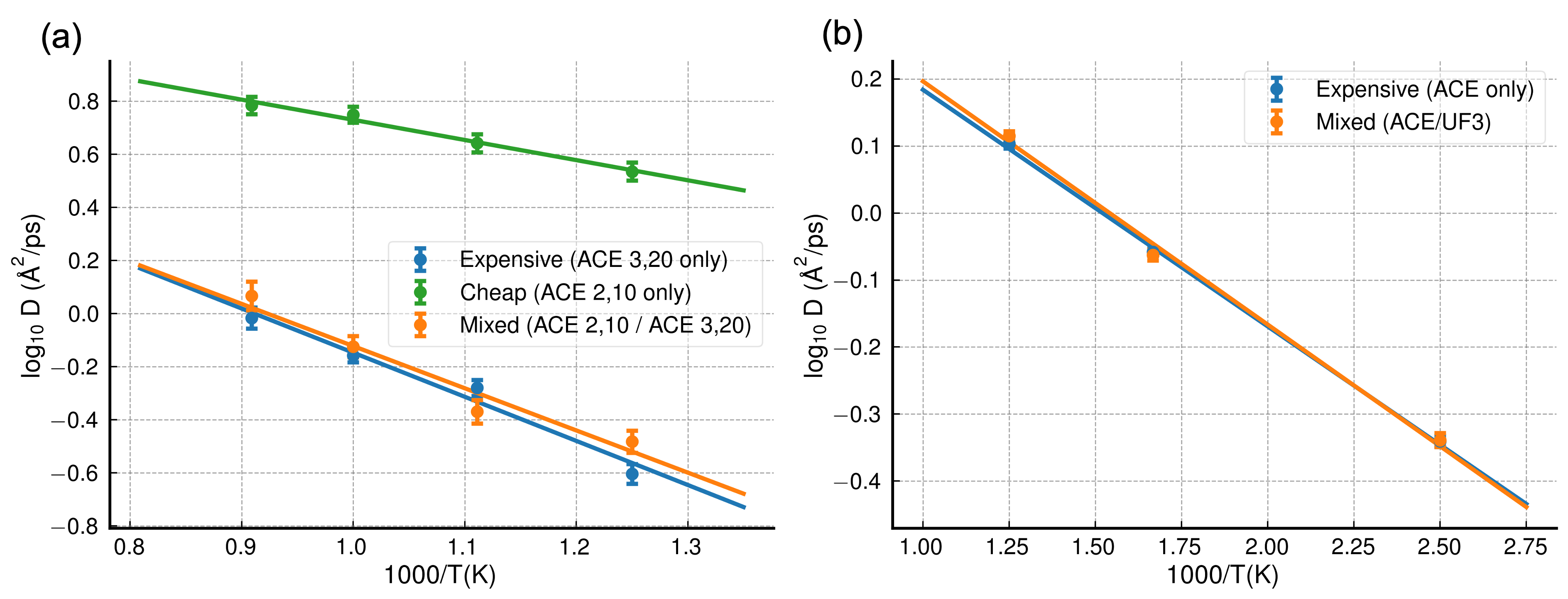}
    \caption{Diffusion coefficient as a function of temperature for the diffusion of the iron interstitial dumbbell defect through bulk iron and for helium in bulk tungsten. \TC{(a)}: The diffusion coefficient of the iron dumbbell for three cases: expensive reference potential only (blue), cheap potential only (orange) and mixed simulation (green). The mixed simulation leads to results which are well matched with the reference. Each point is the average of 5 diffusion coefficients measured in independent runs of 1~ns each. \TC{(b)}: The diffusion coefficient for He in W using a helium-tungsten ACE (blue) and a mixed simulation with both a helium-tungsten ACE and a UF3 tungsten potential (orange). Again, agreement is good. Note that there is no 'cheap-only' reference in this case because the UF3 potential cannot be used to model tungsten-helium interactions. Each point is the average of 25 diffusion coefficients measured in independent runs of 60~ps each. \TC{The error-bars on the points in both (a) and (b) represent the standard error in each value.}}
    \label{fig:Diff_coeff_results}
\end{figure*}

For both target defects, reference simulations were carried out using the expensive ACE potentials. These were followed by ML/ML simulations where the expensive potentials were limited to a small region around the diffusing defect, which was tracked as it moved. Fig.~\ref{fig:Diff_coeff_results} presents the results from these investigations side-by-side. For the Fe system, measurements of the diffusion coefficient as a function of temperature were taken with only the cheap potential, only the expensive potential and using the ML/ML approach. For W-He, only the ML/ML and expensive-only simulation results are presented, as the UF3 potential was fit only for pure W and could not model W-He interactions. For both quantities of interest, the ML/ML simulations agreed well with the all-expensive reference calculations at multiple temperatures, despite the inaccuracy of the all-cheap simulation in the Fe case. The theoretical and measured speedups of the mixed simulations over the reference for both simulation types are shown in Table~\ref{tab:diff_speedup}.

\begin{table}[]
\caption{Speedup table for mixed simulations of diffusion coefficients in Fe and W--He. \TC{The measured serial speedup closely tracks the theoretically predicted value, whilst the parallel speedup is lower due to imperfect load balancing.}}
\label{tab:diff_speedup}
\begin{tabular}{l|lll}
Simulation & \begin{tabular}[c]{@{}l@{}}Theoretical\\ Speedup\end{tabular} & \begin{tabular}[c]{@{}l@{}}Serial\\ Speedup\end{tabular} & \begin{tabular}[c]{@{}l@{}}Parallel\\ Speedup\end{tabular} \\ \hline
Fe & 3.79 & 3.80 & 2.23 \\
W--He & 10.55 & 10.46 & 7.39
\end{tabular}
\end{table}

\rebuttal{\AWE{\subsection{Velocity of screw dislocation in W}}}
\begin{figure*}
    \centering
    \includegraphics[width=0.95\linewidth]{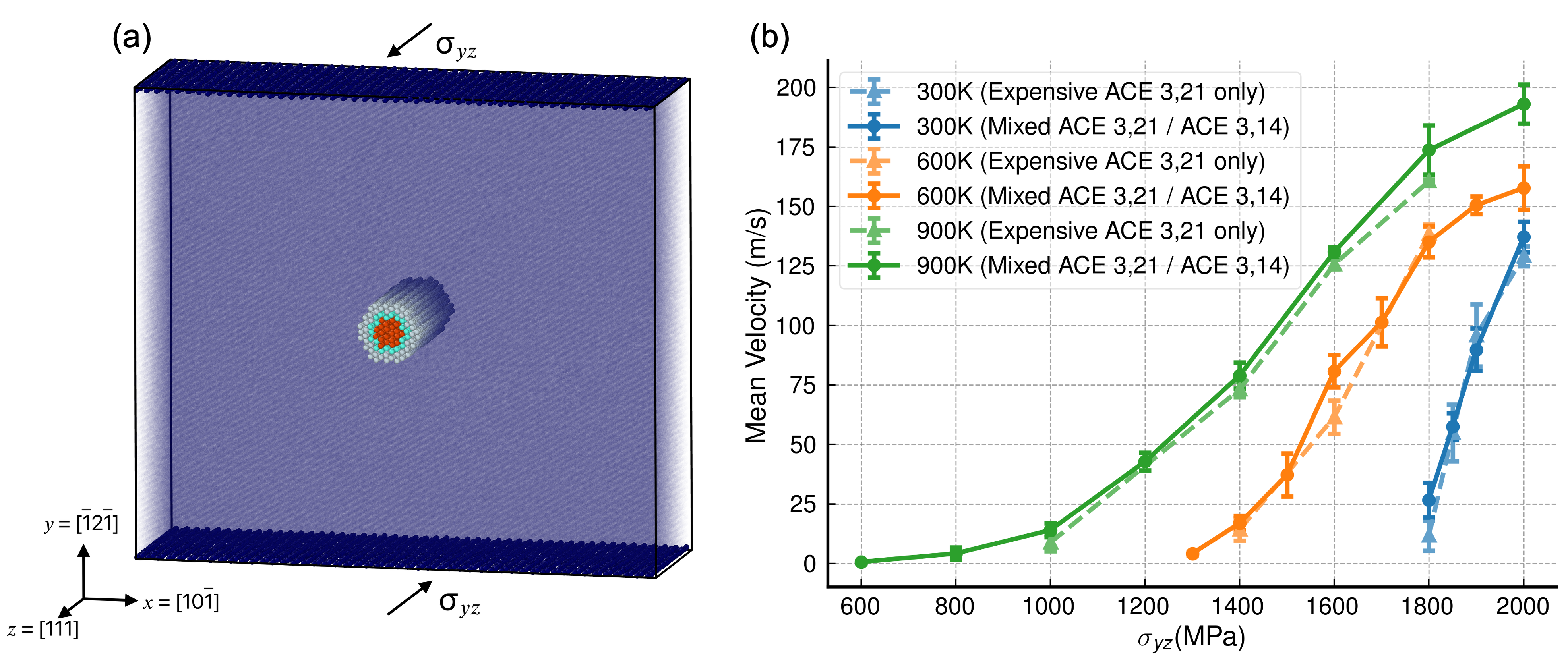}
    \caption{\rebuttal{\AWE{Simulation set-up and results for thermally activated glide of screw dislocations in W at different shear stresses and temperatures. \TC{(a)} Simulation set-up, where dimensions in $x$, $y$ and $z$ are $223~\mathrm{\AA} \times 221~\mathrm{\AA} \times 68~\mathrm{\AA}\ (25~\textit{b})$, respectively, for a total of 195,000 atoms. Stress was applied to the free surfaces at the top and bottom of the cell in $\langle111\rangle$ type directions. The expensive atoms around the dislocation (as detected using common neighbor analysis in LAMMPS) are shown in a column, where the colors correspond to the core (red), blending (green) and buffer (white) regions. \TC{(b)} The velocity results at a range of temperatures and shear stresses. \TC{Green, orange and blue lines represent 300, 600 and 900~K respectively. Results from all-expensive simulations are shown with dashed lines, whilst those from mixed simulations are shown with solid lines. For each point, three separate simulations were conducted and the results were averaged.} Results from the mixed simulations agree well with those obtained from the all-expensive reference. \TC{The error-bars on the points represent the standard error in each value.}}}}
    \label{fig:dislocation_results}
\end{figure*}

\rebuttal{\AWE{The thermally activated glide of screw dislocations is known to play an important role in the deformation of W, with molecular dynamics simulations being a key tool used to understand their mechanics \cite{Bacon_2009}. Following Bacon et al \cite{Bacon_2009} we have used a periodic array of dislocations with Burgers vector $\it{b}$ = $\frac12\langle111\rangle$ to study the thermally activated dislocation glide velocities at a range of applied shear stresses (between 600 and 2000~MPa) and temperatures (300, 600 and 900~K). In each case, all-expensive simulations were conducted using the 3, 21 W ACE potential \cite{Nutter_2024} at a limited range of stresses before ML/ML simulations were conducted using ML-MIX for a greater range, with the expensive region covering only the dislocation.\\}}

\rebuttal{\AWE{The simulation domain is presented in Fig.~\ref{fig:dislocation_results}\TC{(a)}. The ML-MIX expensive region is highlighted. Fig.~\ref{fig:dislocation_results}\TC{(b)} presents the measured velocity-stress profiles obtained in both the all-expensive and mixed simulations. Each data point is the mean of an ensemble of three repeats, with the error bars representing the standard error. The mixed and all-expensive curves agree with each other within error. The all-expensive simulations in this case were approximately $4\times$ the cost of the mixed simulations.}}\\

\rebuttal{\subsection{He implantation into W}}

\rebuttal{Tungsten is a well-known candidate for plasma-facing materials within modern fusion reactors. Proposed as both a diverter and first-wall material \cite{Abernethy_2017,Khripunov_2015}, one of its key properties is the ability to withstand high-energy flux from ions and neutrons. Part of this flux includes impaction from He, the fusion product. In order to predict the damage level and subsequently the longevity of the W material, it is crucial to know (i) how much of the He is expected to enter the W for different incident energies and (ii) how deep the He penetrates. In a previous molecular dynamics study, it was found that these quantities are highly dependent on the choice of interatomic potential \cite{Borovikov_2014}, and so far no reconciliation has been made between molecular dynamics results and the limited experimental data that exists \cite{Borovikov_2014, vanGorkum_1980}}. \\

\begin{figure}
    \centering
    \includegraphics[width=\linewidth]{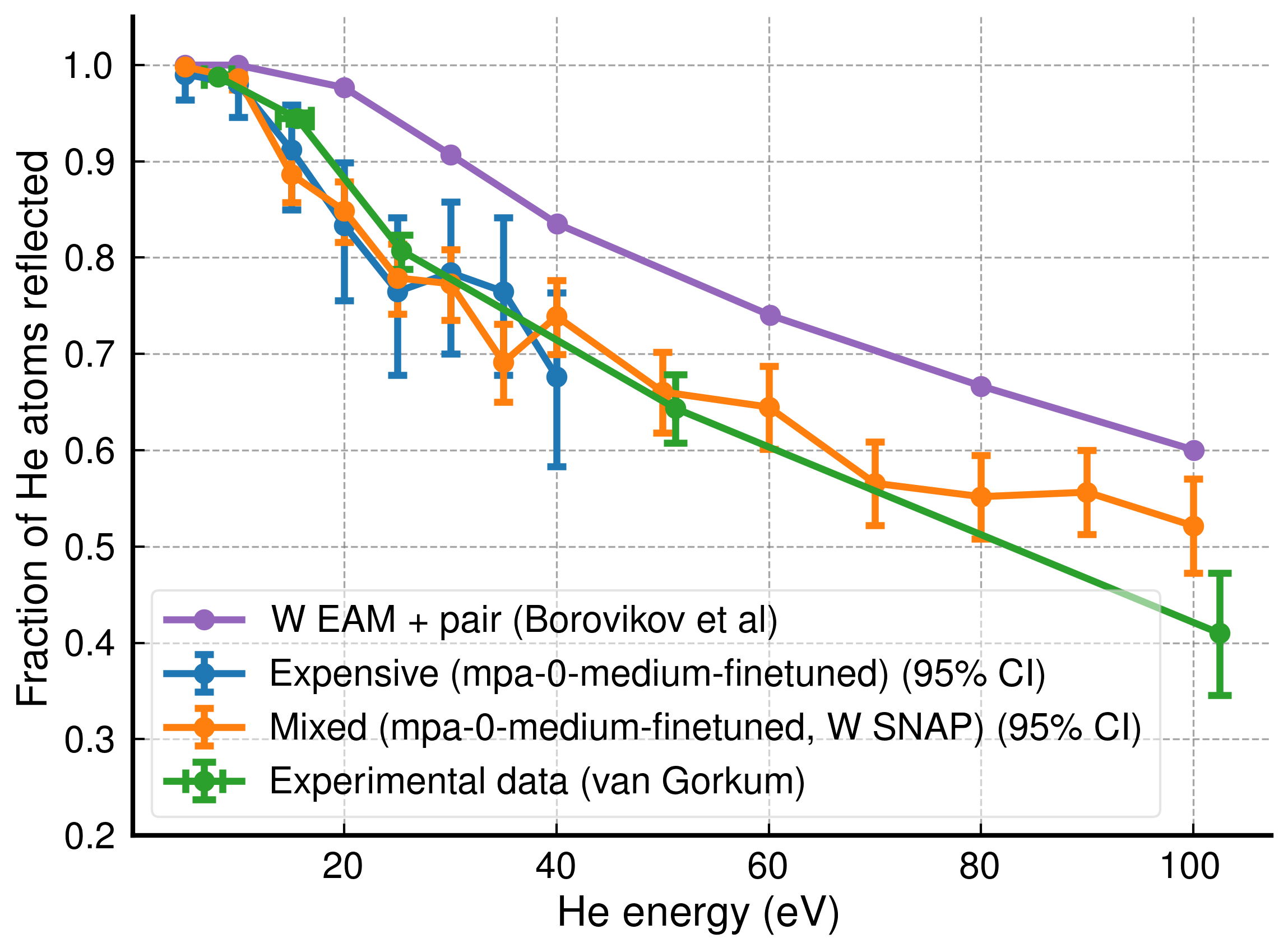}
    \caption{\rebuttal{Reflection coefficient as a function of incident He energy in eV for normal deposition onto a 1000~K W \{100\} W surface. Experimental results from Van Gorkum et al are shown in green \cite{vanGorkum_1980}. Results from the all-expensive potential (fine-tuned mpa-0-medium model) are displayed in blue, and results obtained with ML-MIX are shown in orange. Results from a previous MD simulation using a W EAM and W-He pair potential are displayed in purple \cite{Borovikov_2014}. For the all expensive, to limit computational cost, only energies up to 40~eV were tested with 100 repeats each. With ML-MIX, energies up to 100~eV were repeated 500 times each. Error bars are obtained from the 95\% confidence interval of a fitted Beta distribution.}}
    \label{fig:frac-reflected}
\end{figure}

\begin{figure}
    \centering
    \includegraphics[width=\linewidth]{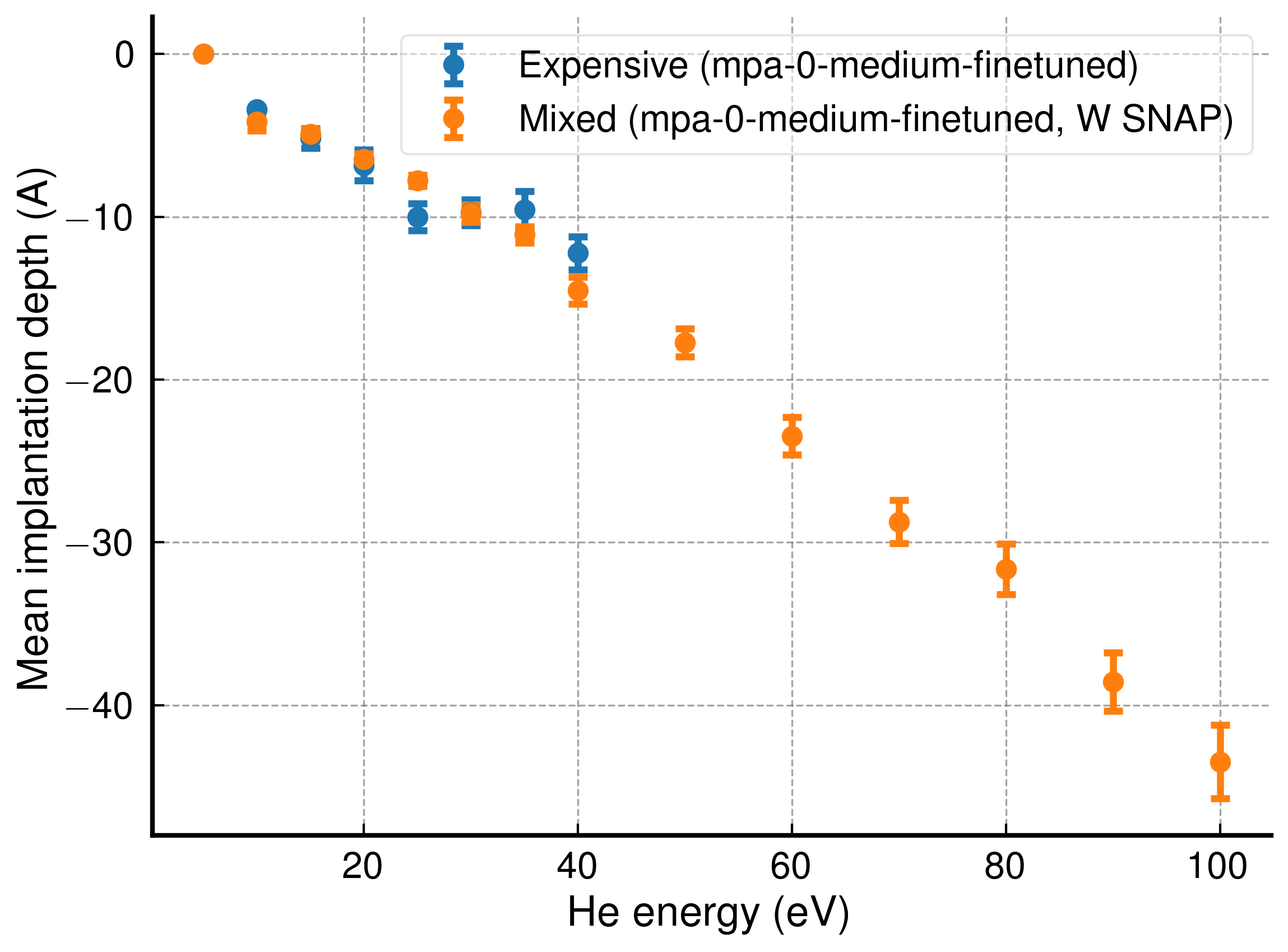}
    \caption{\rebuttal{Implantation depth as a function of incident He energy in eV for normal deposition onto a 1000~K W \{100\} W surface. Results from the all-expensive potential (fine-tuned mpa-0-medium model) are displayed in blue, and results obtained with ML-MIX are shown in orange. For the all expensive, to limit computational cost, only energies up to 40~eV were tested with a 100 repeats each. With ML-MIX, energies up to 100~eV were repeated 500 times each. Error bars represent the standard error in the mean values.}}
    \label{fig:implant-depth}
\end{figure}

\rebuttal{To compare with experiment, we specifically targeted simulations of He implantation into \{100\} W surfaces at 1000~K. Initially, we performed all-expensive simulations using a fine-tuned version of the MACE-mpa-0 foundation model \cite{Batatia_2024}. We then conducted ML/ML simulations using in which the expensive potential interactions were limited to the immediate vicinity of the He atom. In both all-expensive and ML/ML simulations, all computation was conducted on GPU. Fig.~\ref{fig:frac-reflected} shows the measured reflection coefficient as a function of incident He energy. The all-expensive (mpa-0-medium-finetuned) data (blue points), ML-MIX data (orange points) and experimental data (green points) all broadly agree within error. Average implantation depths were also measured for each energy. These are shown in Fig.~\ref{fig:implant-depth}, where it can be seen that results from the ML/ML simulation are in broadly good agreement with the all-expensive. The reduced cost of the ML/ML simulations allowed extension to larger system sizes and thus greater He energies and implantation depths.} \\

\rebuttal{The speedup of the mixed simulation over the expensive simulation was approximately 10$\times$ for simulations where a reflection takes place and approximately 5$\times$ for simulations where the He implanted.}\\

\section{Discussion}
\label{sec:Discussion}
We have presented ML-MIX, a hybrid simulation package which aims to break existing cost-accuracy tradeoffs in simulations by localizing accuracy requirements, enabling simulation speed-ups of $4-10 \times$ with minimal loss in accuracy on quantities of interest. \rebuttal{ML-MIX has the following advantages that set it apart from previously published similar ML/MM schemes: (i) ML-MIX is an easy to install, direct plugin to \texttt{LAMMPS}, which is fast and has no external dependencies, and aims to be compatible with a large number of existing pair\_styles. (ii) ML-MIX functions as an external wrapper around \texttt{LAMMPS} pair\_styles, meaning there is no necessary editing of C++ code, (iii) ML-MIX works with existing \texttt{LAMMPS} parallelism and load-balancing, and (iv) ML-MIX (though our Kokkos implementation) is compatible with pair\_styles that execute on both CPU and GPU}. \\

We have described two ways in which cheap potentials can be generated; either by fitting directly to the same underlying DFT data as the expensive reference, or by fitting to synthetic data generated by the expensive potential \TC{in a targeted region of the potential energy surface}. \TC{We found that fitting to synthetic data was particularly important for very lightweight potentials, as for these we observed that attempting to fit to the diverse reference DFT data directly lead to unacceptably high RMSEs.} To ensure close matching of specific properties, we developed a constrained fitting scheme, the details of which are described in section~\ref{sec:constrained_fit}. For linear ACE potentials, we found that imposing constraints allowed for exact matching of elastic constants between the cheap and expensive potentials. For UF3 potentials, a less flexible architecture, we have shown that under hard constraint elastic constants improve up to some limit, beyond which any further improvements were limited by the misspecification of the functional form. We stress that whilst in this study we have specifically discussed constrained fitting in the context of distilling one MLIP from another, it would be easy to apply this methodology to constrain the fitting of a large, flexible MLIP to DFT data directly, extending recent retraining approaches \cite{Grigorev_2023}. \\

We found that constrained fitting is particularly important for small potentials of $<50$ basis functions; without it, errors are far more pronounced at small strains. This can lead to erroneous Eshelby-inclusion \cite{eshelby1957} like strain fields in the vicinity of the ML/ML potential boundary under small deformations. This is explored in more detail in section S3 of the supplementary information.\\

The first case-study we present in this paper was a nudged elastic band energy barrier calculation for the migration of an Si vacancy. We showed that, using an ML/ML simulation, a value for the migration barrier can be attained which is within 1~meV of the all-expensive reference at a fifth of the computational cost. This speedup, whilst already considerable, is not as great as what is seen for ML/ML MD simulations on the same 8000 atom domain -- this is because a larger fixed set of expensive atoms is necessary for a NEB to converge. It is important to point out that total energies are undefined in ML-MIX ML/ML simulations; the final energies were computed by performing one-shot energy evaluations of the relaxed structures using only the expensive potential.\\

Next, we presented an investigation into the average force on a stretched and rigidly fixed bond in an Si bulk structure. This was selected to mirror previous tests performed in QM/MM studies \cite{Bernstein_2009}. It should be noted that for an $r_{\mathrm{core}}$ of 0~{\AA} (cheap potential only) the average force measured on the stretched bond is within 5\% of the correct value, which is a significantly smaller error than one typically sees in QM/MM simulations. Adding even a small amount of expensive potential region around the stretched bond led to an immediate improvement in the measured average bond force, leading to a value which was within $1$ standard error of the $NVE$ reference value. This behavior persisted for larger $r_{\mathrm{core}}$ values.
%This very small error also persisted for larger $R_{\mathrm{core}}$ values. 
The measured serial speed-up closely matched theoretical predictions, showing that our implementation incurs minimal overhead due to the mixing process. 
%is added to the simulation in the potential mixing and region rebuilding process. 
However, in parallel testing (on 48 cores) the measured speedup is below  theoretical predictions, in particular for the $r_{\mathrm{core}}=4\ \mathrm{\AA}$ case. 
We attribute this performance to the limited strong scalability of the 
expensive interatomic potential at low atom count, leading to significant 
load balancing issues in \texttt{LAMMPS} when there are few expensive atoms. The limited speed-up is thus a property of the expensive interatomic potential rather than our load balancing protocol; near-theoretical parallel speed-ups are achievable on larger systems with simple load balancing requirements. This is demonstrated in section S2 of the supplementary information. \\

We also presented case studies on defect diffusion coefficients in Fe and W. The measured diffusion coefficients consistently agreed with the all-expensive reference simulations within statistical error. The close matching of the measured diffusion coefficients with the reference values suggest that one only requires spatially local accuracy to attain correct defect diffusion dynamics in materials; corroborating findings from earlier QM/MM studies \cite{Csányi_2004}. \\

\rebuttal{Next, we considered some more complex and scientifically relevant case-studies, aiming to demonstrate the type of simulations where we believe ML-MIX will find application. These were (i) a study of the thermal glide of screw dislocations in W at a range of temperatures and pressures, and (ii) a study of normal deposition of He in W at 1000~K.} \\

\rebuttal{\AWE{We considered the thermally activated glide of $\it{b}$ = $\frac12\langle111\rangle$ screw dislocations in W. We found that in a mixed ACE/ ACE simulation, we were able to reproduce the dislocation velocities measured in the all-expensive reference, meaning that the mixed simulation was able to reproduce the correct dynamics of the kink-nucleation-propagation mechanism that controls the glide of screw dislocations in W. Whilst there was scope to study longer dislocation lines, we found increasing the length of the dislocation increased the probability of cross-kinking -- a phenomenon that can lead to debris formation. This can cause nonphysical interference with subsequent dislocation motion when the dislocation re-encounters the debris due to periodicity. As such, we leave an exploration of larger dislocation studies with ML-MIX to future work.}} \\

%TODO update energy with final simulation result
\rebuttal{Finally, we simulated the normal implantation of He into (100) W surfaces at He energies up to 70~eV at 1000~K. This led to a novel result; for the first time, experimental observations of reflection coefficient were reproduced up to 40~eV with the mpa-0-medium foundation model fine-tuned to W--He DFT data. Following this, we carried out the same simulations with ML-MIX, where we limited interactions with the expensive potential to directly around the He and used a pure W SNAP potential for the remainder of the interactions. In order to ensure efficient evaluation of the expensive MACE potential, ML-MIX was ported to GPU using Kokkos \cite{kokkos1, kokkos2}. Using ML-MIX, we showed that (i) up to 40~eV results agreed well with the all expensive simulation and experiment, and (ii) above 40~eV results continue to line up closely with experiment, with deviations only occuring at relatively high energy ($>80$~eV). Measurements of implantation depth were also taken in each case, and those from the mixed simulations aligned well with those from the all-expensive simulations. Due to the relative low cost of the mixed simulations, we were able to perform more repeats and create the larger cells necessary for high energy simulations without needing to parallelize over more GPUs.} \\

\rebuttal{It is important to note that the experiment we have drawn comparisons with involved incident He$^{+}$ ions \cite{vanGorkum_1980}, whilst in this study we are considering incident He atoms. One possible explanation for our close agreement despite this is that in reality, as the He$^{+}$ approaches the surface, it gains an electron from the W conduction band on timescales faster than the movement of the nuclei. A recent time-dependent DFT study by Ito et al \cite{Ito_2025} studying plasma-wall interactions observed electron transfer in the case of He$^{2+}$, but was inconclusive on He$^{+}$. We believe that reconciling this and performing a wider exploration of the implantation parameter space are both beyond the scope of this primarily method-focused paper, and therefore leave them as interesting directions for future work.} \\

\rebuttal{A key limitation in ML/ML (and indeed, QM/MM) simulations arises from the lack of energy conservation; if a force mixing method is used to combine potentials (as it is here), simulations do not conserve energy. This means that a thermostat may be required to keep simulations from substantially heating up as they progress. The energy drift in well-matched ML/ML simulations is relatively small ($\mathcal{O}(10~\mathrm{K/ps})$), meaning thermostats can be weak enough to not impact dynamics --- but this should \textbf{always be checked}. Force mixing is discussed further in section \ref{sec:force_mixing}, and a full investigation into the factors affecting energy conservation in ML/ML simulations is presented in section S4 of the supplementary information.} \\

\rebuttal{A possible solution to this problem is the use of so-called `energy-mixing' \cite{Bernstein_2009}, where a Hamiltonian is constructed for the mixed system, for example by writing the total energy as a sum of local site energies from each potential. Upon testing this scheme however, we found the forces derived had errors of $>50$\% in the vicinity of the ML/ML boundary, even when using potentials that nominally are well matched in force-mixing. This backs up previous work on energy-mixing for QM/MM conducted by Bernstein et al \cite{Bernstein_2009} and expanded on by Chen \& Ortner \cite{Chen_Ortner_2017} where it was found that eliminating these spurious `ghost forces' in energy-mixed QM/MM for infinitesimal displacements away from equilibrium required the MM potential have an exactly matching force-constant matrix (second order derivatives of the site energies with respect to atomic positions). This requirement is stringent and as such would require more flexible (and thus expensive) cheap potentials, and even if satisfied only would eliminate ghost forces for atoms near equilibrium. Additionally, if the regions change over time as atoms move (a common feature of long-timescale ML/ML simulations), energy would not be conserved anyway; as regions update, potential energy would be continually added or removed from atoms. Conversely, a buffered force-mixing approach naturally has no spurious ghost-forces and can be made to work with far simpler cheap potentials. In view of this, we believe (despite the lack of energy conservation) that force-mixing is the clear choice for ML/ML simulations.} \\

\rebuttal{A further limitation is that we have made no attempt to quantify or estimate how the error associated with the size of the expensive region selected may propagate through to quantities of interest. Uncertainty quantification (UQ) for MLIPs and QM/MM is a very active area of research \cite{Swinburne_2025, Perez_2025, Best_2024, Wang_2024}, and we believe that applications of UQ to ML/ML simulations is a very exciting possible direction for future work.} \\

Through these studies, we have demonstrated the flexibility and efficiency of the ML-MIX package. In our initial, simple case-studies, we show that ML-MIX can be used to accelerate simulations without compromising accuracy both in simple cases where the same atoms are always involved in the defects (i.e, for He in W) and for more complex cases where different atoms are involved as the defects move (i.e, for Fe dumbbell interstitials in Fe). \rebuttal{In two further case-studies, we have shown that ML-MIX can be used to attain a significant speedup without any meaningful loss in accuracy in applications of genuine scientific interest. In the first of these, we present results for the speed of $\it{b}$ = $\frac12\langle111\rangle$ screw dislocations in W at a range of temperatures and shear stresses. In the second, we provide estimates of reflection coefficients and implantation depths for normally deposited He into 1000~K \{100\} W surfaces. Our measured reflection coefficients match experimental observations up to a deposition energy of 80 eV.} Looking forwards, we anticipate that ML-MIX will find application in a broad range of atomic simulations of heterogeneous systems, including the calculation of defect free energy barriers \cite{Swinburne_2018}, interface migration \cite{Rogal_2021} and nucleation phenomena \cite{Bonati_2021}. 

\section{Methods}\label{sec2}
\subsection{Constrained potential fitting}
\label{sec:constrained_fit}
\begin{figure}
    \centering
    \includegraphics[width=0.8\linewidth]{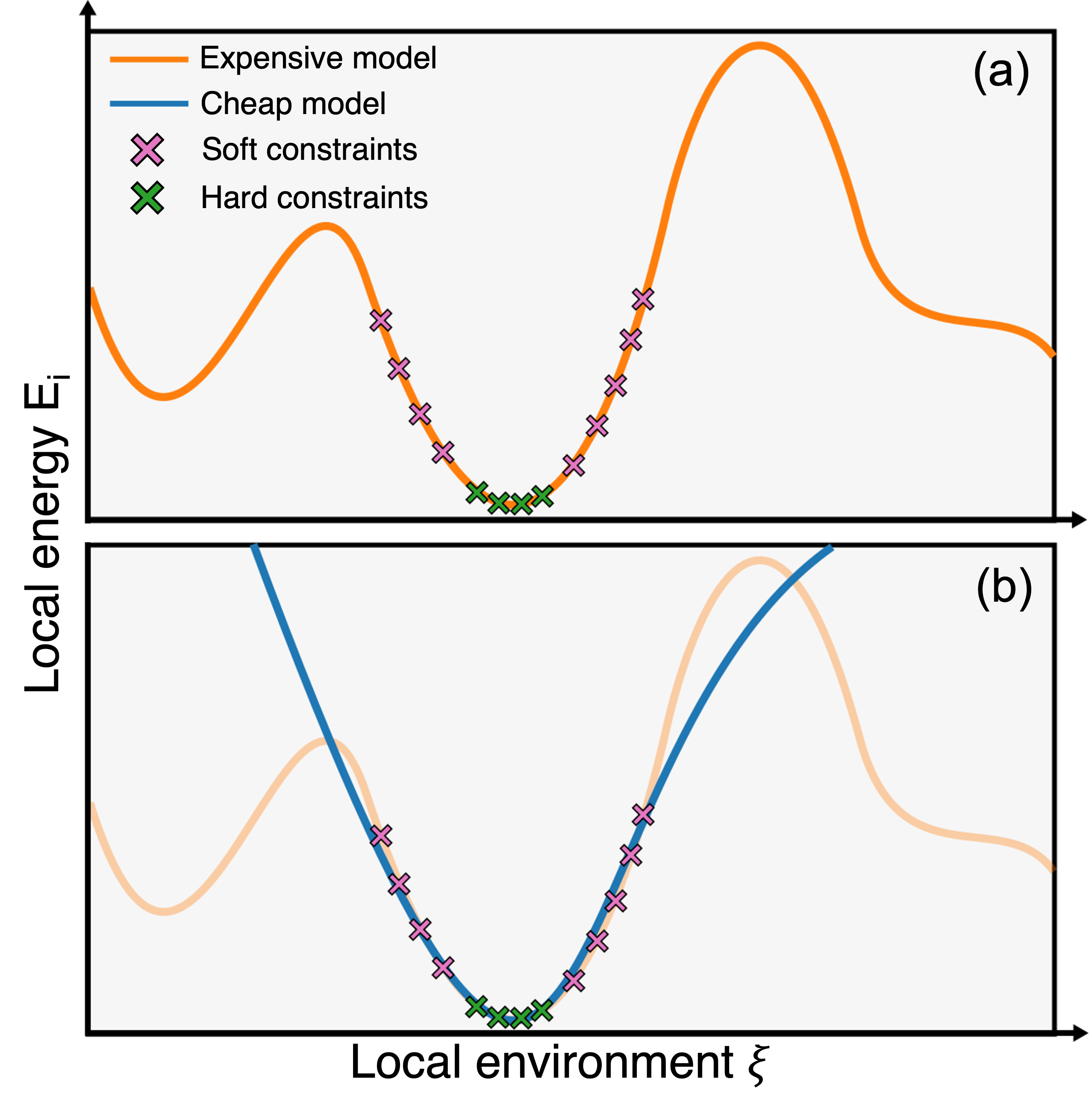}
    \caption{A schematic representation of the constrained fitting of a cheap potential to an expensive one. \TC{(a)}: A representation of the local potential energy surface of the expensive reference model (orange curve). Data has been sampled from this potential energy surface (crosses). \TC{(b)}: The cheap model, fit to this data in order to locally match the expensive model. It approximately matches the `soft' constraint data (pink crosses) and exactly matches the `hard' constraint data (green crosses).}
    \label{fig:local_fitting}
\end{figure}

The constrained potential fitting process is shown schematically in Fig~\ref{fig:local_fitting}. By fitting a cheap potential to highly localized regions of the potential energy surface of the expensive potential, it is possible to attain high accuracy in local regions whilst maintaining a low complexity (and thus low evaluation cost).  \\

% \subsubsection{Selecting constraints}

\begin{figure}
    \centering
    \includegraphics[width=0.95\linewidth]{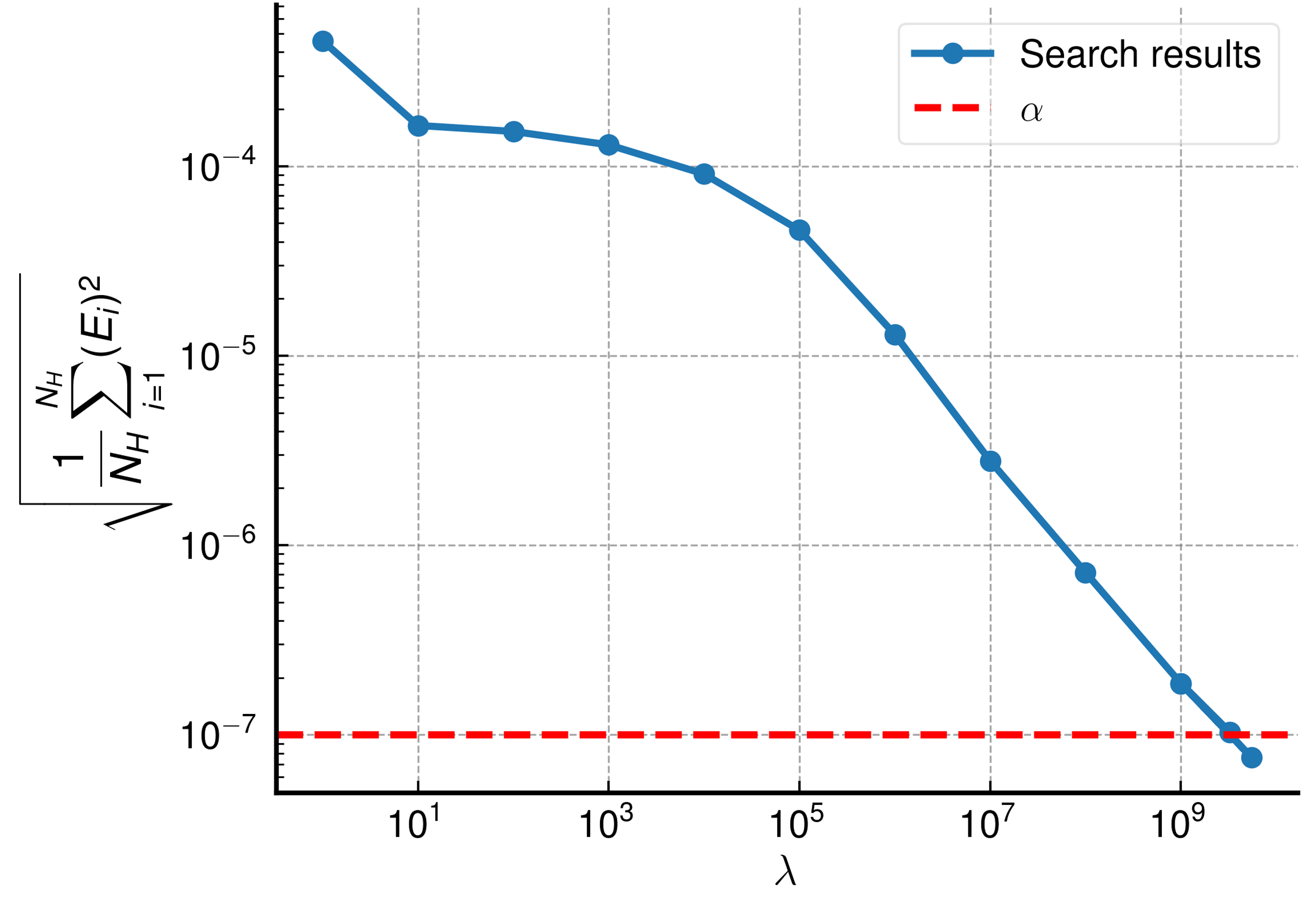}
    \caption{\rebuttal{Automated selection of lambda value during a single constrained fit to satisfy a given alpha. The vertical axis shows the energy RMSE per atom. The target alpha value is shown as a dashed red horizontal line. \TC{Each blue point represents a step on the search.}}}
    \label{fig:lambda_search}
\end{figure}

\TC{We first consider the challenge of selecting suitable constraints.} The requirements of the cheap potential are split into two types: soft constraints (loose matching), and hard constraints (tight matching). Archetypal hard constraints are the elastic constants, to ensure seamless matching of the long range elastic stress fields between the cheap and expensive potentials in a solid state simulation~\cite{Bernstein_2009}. \\

Writing the design matrices for the hard and soft constraints as $\mathbf{A}_\mathrm{H} \in \mathbb{R}^{N_{\mathrm{H}} \times N_{\mathrm{D}}}$ and $\mathbf{A}_\mathrm{S} \in \mathbb{R}^{N_{\mathrm{S}} \times N_{\mathrm{D}}}$, and the fitting data as $\mathbf{y}_{\mathrm{H}} \in \mathbb{R}^{N_{\mathrm{H}}}$ and $\mathbf{y}_{\mathrm{S}} \in \mathbb{R}^{N_{\mathrm{S}}}$, we can express the potential parameters $\mathbf{c} \in \mathbb{R}^{N_{\mathrm{D}}}$ as a solution to the constrained optimization problem
\begin{equation}
    \min_{\mathbf{c}, ||\mathbf{A}_\mathrm{H} \mathbf{c} - \mathbf{y}_{\mathrm{H}}||^{2} < \alpha} \left(||\mathbf{y}_{\mathrm{S}} - \mathbf{A}_\mathrm{S}\mathbf{c}||^{2} \right)
\end{equation}
where regularization terms have been excluded for simplicity. The hard constraint is imposed by constraining the fit to the subspace $||\mathbf{A}_\mathrm{H} \mathbf{c} - \mathbf{y}_{\mathrm{H}}||^{2} < \alpha$, where $\alpha$ is the maximum allowed error for the model on the hard constraint data. Expressing this constraint through a Lagrange multiplier argument gives us the augmented Lagrangian
\begin{equation}
    \mathcal{L}(\mathbf{c}, \lambda) = ||\mathbf{y}_{\mathrm{S}} - \mathbf{A}_\mathrm{S}\mathbf{c}||^{2} + \lambda (||\mathbf{y}_{\mathrm{H}} - \mathbf{A}_\mathrm{H} \mathbf{c}||^{2} - \alpha).
\end{equation}
By requiring that $\nabla_{\mathbf{c}}\mathcal{L}(\mathbf{c}, \lambda) = 0$, we can express optimal constrained model parameters $\mathbf{c}$ at fixed $\lambda$ as a solution to the unconstrained problem
\begin{equation}
    \label{eq:full_loss}
    \min_{\mathbf{c}}(||\mathbf{y}_{\mathrm{S}} - \mathbf{A}_\mathrm{S}\mathbf{c}||^{2} + \lambda ||\mathbf{y}_{\mathrm{H}} - \mathbf{A}_\mathrm{H} \mathbf{c}||^{2}).
\end{equation}
From this expression, it is clear that the quadratic constraint is equivalent to adding the hard constraint configurations to the fit with weights scaled by $\lambda$. Obtaining a solution to the full constrained problem therefore requires finding the minimum $\lambda$ such that the solution to (\ref{eq:full_loss}) lies on the surface of the constraint subspace $||\mathbf{A}_\mathrm{H} \mathbf{c} - \mathbf{y}_{\mathrm{H}}||^{2} = \alpha$. \\

Scripts to perform constrained linear fitting for ACE and UF3 potentials are provided in the accompanying code \cite{ML-MIX}. To perform the $\lambda$ hyperparameter search, the linear problem is solved repeatedly. Initially, $\lambda$ is increased logarithmically until $||\mathbf{A}_\mathrm{H} \mathbf{c} - \mathbf{y}_{\mathrm{H}}||^{2} < \alpha$. Interval bisection is then used to find $\lambda$ such that $||\mathbf{A}_\mathrm{H} \mathbf{c} - \mathbf{y}_{\mathrm{H}}||^{2} - \alpha < \mathrm{tol}$, where $\mathrm{tol}$ is set by default as $\alpha/10$. \rebuttal{An example $\lambda$ search to satisfy a constraint of $\alpha = 10^{-7}$ is shown in Fig.~\ref{fig:lambda_search}}. $\mathbf{A}_\mathrm{H}$ and $\mathbf{A}_\mathrm{S}$ only need to be assembled once, and as the cheap potentials are small (often $\lesssim 100$ parameters), the whole constrained fitting process takes approximately 5 minutes on a single CPU core. \\

\subsection{Generation of synthetic data}

\begin{figure*}
    \centering
    \includegraphics[width=0.95\linewidth]{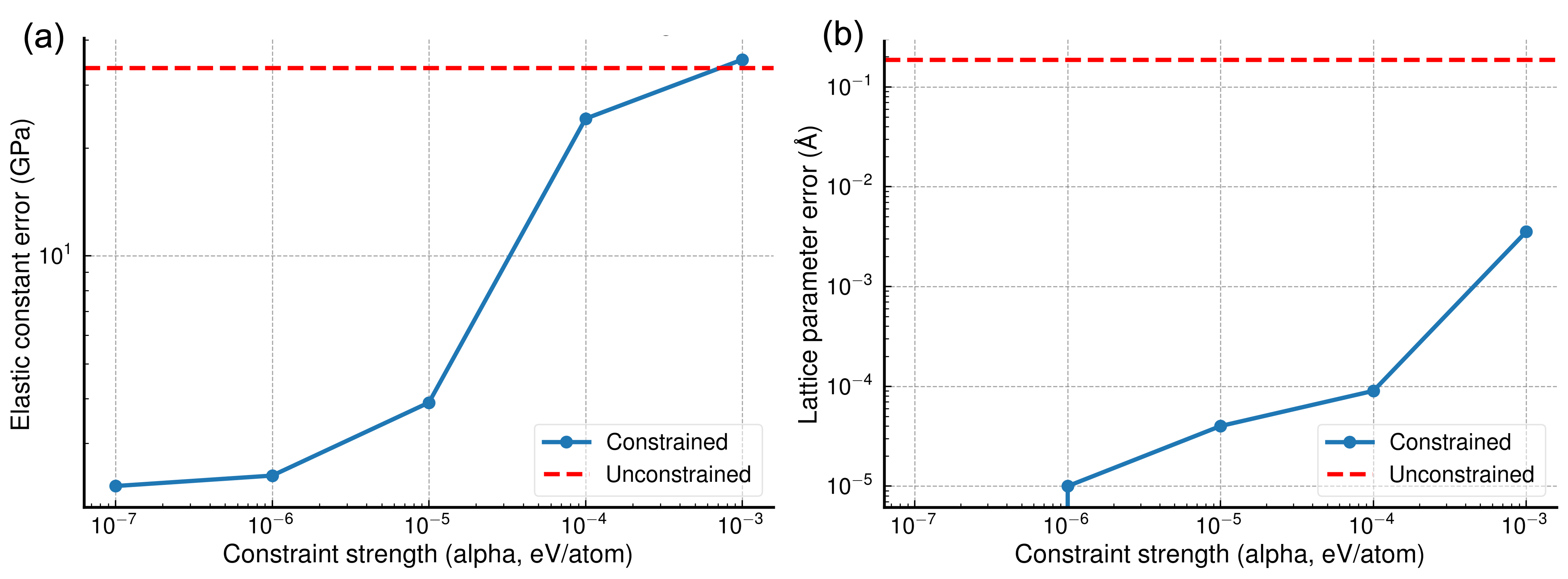}
    \caption{\rebuttal{Demonstration of convergence of key quantities with hard constraint strength for the Si ACE 2, 10 cheap model. Error is shown compared to the Si ACE 4, 20 expensive model. In each case, the unconstrained case is shown by a dashed horizontal red line \TC{and the result at each constraint strength ($\alpha$) is shown by a blue point}. \TC{(a)}: The L2 norm of the error over $\mathrm{C}_{11}$, $\mathrm{C}_{12}$ and $\mathrm{C}_{44}$. \TC{(b)}: The absolute error in lattice parameter. Note that at a constraint strength of $10^{-7}$, the lattice parameter error was 0 up to the number of decimal places measured (5).}}
    \label{fig:combined_constraint_conv}
\end{figure*}

Simple high-temperature bulk equilibrium dynamics was selected as the soft constraint data. For Si, a 4096-atom diamond structure bulk cell was simulated at 500~K for 10~ps, with snapshots saved every 0.5~ps. For W and Fe, 2000-atom body centered cubic (BCC) bulk cells were simulated at 1200~K, with snapshots saved every 0.1~ps. The soft constraints consisted of the total energy and forces from each snapshot. \\

For the hard constraints, following \cite{Grigorev_2023}, data was generated by applying multiple shear and expansion deformations to  $5\times5\times5$ bulk supercell of Si, Fe and W lattices and saving the resulting energies. 17 configurations were generated between $\pm 0.5 \% $ strain for all of the following strain states: hydrostatic, uniaxial [100], uniaxial [111], shear ([100], (010)), shear ([110], (001)) and shear ([110], (1$\Bar{1}$2)). As the energy of weak, homogeneously deformed lattice structures are described fully by the elastic constants \cite{Grigorev_2023}, fitting to the selected hard constraint enforced elastic constant matching between cheap and expensive potentials. \\

For the Si and Fe ACE potentials, an $\alpha$ value was selected that corresponded to an enforced energy error of less than $10^{-7}$~eV/atom on the hard constraint data. For Si, the corresponding $\lambda$ value was $3.25 \times 10^{9}$. For Fe, $\lambda$ was found to be $5.5 \times 10^{9}$. To a reader familiar with MLIPs, a target per-atom-energy error of $\alpha = 10^{-7}$~eV/atom may seem very small (a standard aim when fitting MLIPs is a per-atom-energy error of around 1~meV). However, we found that pushing $\alpha$ to such small values was necessary to achieve the tight property matching that we were aiming for. This is demonstrated for the Si cheap potential in Fig.~\ref{fig:combined_constraint_conv}; such small $\alpha$ values were necessary to match elastic constants within a fitting error of $\sim 2$~GPa and lattice parameter within a fitting error of $10^{-5}$~{\AA}. For the W potential the constrained fitting process failed due to the UF3 potential not being capable of simultaneously fitting the hard constraint data and maintain a low RMSE on the soft constraints. By searching the space of $\lambda$ values, it was found that above approximately $\lambda$ = $4.3 \times 10^{5}$, $||\mathbf{A}_\mathrm{H} \mathbf{c} - \mathbf{y}_{\mathrm{H}}||^{2}$ stopped decreasing. $\lambda$ was therefore set to $4.3 \times 10^{5}$. \\

\rebuttal{\subsection{Additional DFT data}}
%Section plan - two cheap potentials were additionally fit unconstrained
\rebuttal{In the case of the cheap and expensive potential fit for the W dislocation and He -- W implantation studies, additional DFT data was acquired. The DFT parameters used matched those described in \cite{Nutter_2024_supp}. \AWE{For the dislocation mobility simulations, the dataset in \cite{Nutter_2024_supp} was extended to include 1000~K surface and dislocation kink configurations. 40 structures for both {(110)} and {(112)} surfaces were generated, each containing 45 and 48 atoms, respectively. 12 structures for each of the two kink types (left and right) were generated, containing 330 and 345 atoms, respectively}. For the He -- W implantation study, 129 snapshots of small 91-atom implantation cells were generated, from simulations of He implantations at 10~eV and 100~eV. These were added to the full He -- W dataset provided by Nutter et al \cite{Nutter_2024}, which was then used for generation of the expensive and cheap potentials for the He -- W implantation simulation.}\\

\subsection{Region tracking}

\begin{figure*}
    \centering
    \includegraphics[width=0.9\linewidth]{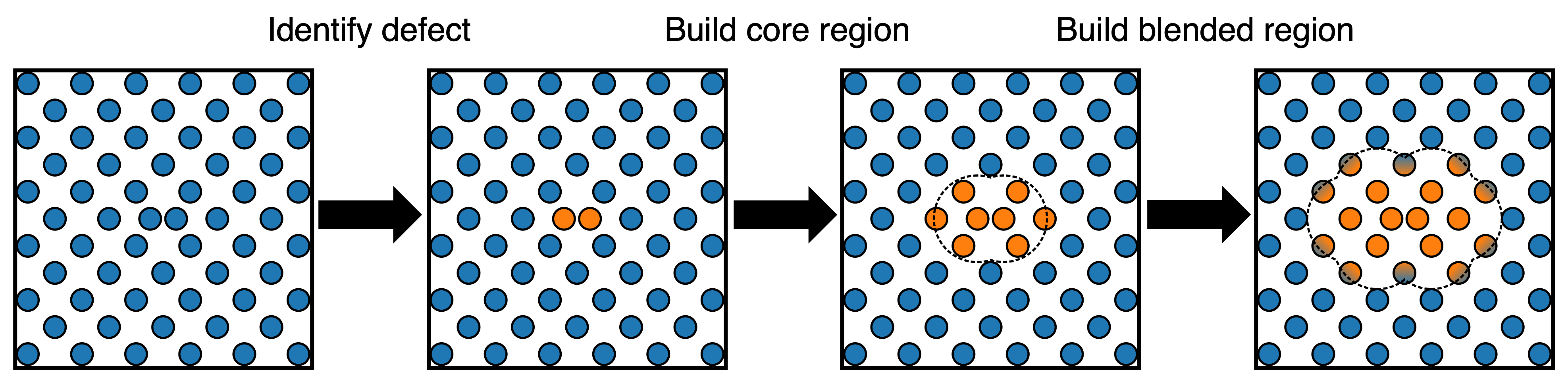}
    \caption{Region construction schematic. From left to right: Starting with a cell containing an initial defect, it is identified (orange atoms in panel two) either using a user-specified \texttt{LAMMPS} group or by the output vector from a user-defined \texttt{LAMMPS} fix. The defect atoms are then iterated over, with atoms that sit less than one core radius out from any defect atom being added to the core region. This construction process is then repeated for the atoms in the blended region.}
    \label{fig:TrackingFigure}
\end{figure*}

When running an ML/ML simulation, it is crucial that process of tracking defects and building the expensive potential region(s) is: \\
(i) Accurate --- if the cheap potential is erroneously used in a region that requires the expensive potential, it could impact on the accuracy of the simulation. \\
(ii) Flexible --- any defect type should be trackable. \\
(iii) Fast --- region building should not add considerable overhead to a simulation. \\

In order to satisfy all three of these simultaneously, the region building and tracking were implemented directly in \texttt{LAMMPS} through a custom fix. By re-using the same neighbor lists that are built for the pair style evaluation step, this fix can quickly rebuild regions. Additionally, the fix is MPI local; whilst a limited amount of communication between neighboring processes is necessary, no expensive global communication is required. \\

The fix builds regions in two distinct stages: (i) Identify the `seed' atoms through user-defined criteria. (ii) Construct the regions around these atoms.

% \subsubsection{Seed atom identification}
Seed atoms are identified in one of two ways; through a \texttt{LAMMPS} group or through querying the output vector of a separately defined \texttt{LAMMPS} fix. Tracking through a predefined \texttt{LAMMPS} group is for situations where seed atoms are unchanging, and are known at the start of the simulation, as is the case with He atoms in W. Tracking through the output vector of a user-supplied fix is useful for tracking defects which are not tied to any one particular set of atoms, for example a vacancy, dislocation or an Fe dumbbell interstitial. Seed atom identification is schematically represented in the first two panels of Fig.~\ref{fig:TrackingFigure}.

% \subsubsection{Region construction}
Once the seed atoms are identified, regions are constructed around these seed atoms, using a method that is schematically identified in the last two panels of Fig.~\ref{fig:TrackingFigure}. \\

During region construction two arrays are populated, \texttt{i2\_potential} and \texttt{d2\_eval}. Both arrays are of size number of potentials $\times$ number of atoms. The \texttt{i2\_potential} array holds integers and is used to determine which atoms are to be evaluated with each potential; \texttt{i2\_potential[1][10] = 1} indicates that atom 10 should be evaluated with potential 1, whilst \texttt{i2\_potential[1][10] = 0} indicates that it should not be. \texttt{d2\_eval} holds double-precision real numbers and describes how much each potential contributes to forces on atoms. By allowing \texttt{d2\_eval} to hold fractional values, it is possible to introduce a blending region in which atoms move under a mixture of the forces from the two separate potentials; \texttt{d2\_eval[0][10] = 0.4} and \texttt{d2\_eval[1][10] = 0.6} indicates that atom 10 should get 40\% of its' forces from potential 0 and 60\% from potential 1. \rebuttal{Two possible blending functions are implemented in ML-MIX - linear blending and cubic blending. A brief discussion of the impact on energy conservation of these two blending schemes is presented in section S4 of the supplementary information}. \\

Three regions are built in the region construction process, the \textbf{core region}, in which the force on all the atoms is given entirely by the expensive potential, the \textbf{blending region}, in which the forces are a mixture between the cheap and expensive potentials and the \textbf{buffer region}, which is crucial for the accuracy force-mixing scheme, and is discussed in the next section. The size of each region is specified by user controlled parameters: \texttt{r\_core}, \texttt{r\_blend} and \texttt{r\_buffer} which are passed to the fix. \\

Algorithms \ref{alg:build_core_region}, \ref{alg:build_blend_region} and \ref{alg:build_buffer_region} describe how these arrays are populated in the region construction process for one of the two potentials (potential 0). Note that in algorithm \ref{alg:build_blend_region}, a linear blending function to mix together potentials in the blending region is used. This is not mandatory and in principle could be replaced by any blending function. So long as there are only two potentials, it is trivial to populate \texttt{i2\_potential} and \texttt{d2\_eval} for the other potential once the regions have been constructed around the seed atoms. Currently, ML-MIX is limited to two potentials only. \\

\begin{algorithm}
\caption{Build core region}
\label{alg:build_core_region}
\begin{algorithmic}[1]
    \State Initialise i2\_potential to 0
    \State Initialise d2\_eval to 0.0
    \For{i $\leftarrow$ seed atoms}
        \State i2\_potential[i][0] = 1
        \State d2\_eval[i][0] = 1.0
        \For{j $\leftarrow$ neighbors of atom i}
            \If{$||\mathrm{pos(i)} - \mathrm{pos(j)}||$ $<$ r\_core}
                \State i2\_potential[j][0] = 1
                \State d2\_eval[j][0] = 1.0
            \EndIf
        \EndFor
    \EndFor
    \State Communicate()
\end{algorithmic}
\end{algorithm}

\begin{algorithm}
\caption{Build blended region}
\label{alg:build_blend_region}
\begin{algorithmic}[1]
    \For{i $\leftarrow$ core atoms}
        \For{j $\leftarrow$ neighbors of atom i}
            \If{$||\mathrm{pos(i)} - \mathrm{pos(j)}||$ $<$ r\_blend}
                \State i2\_potential[j][0] = 1
                \State r = $||\mathrm{pos(i)} - \mathrm{pos(j)}||$
                \State v = max(d2\_eval[j][0], 1.0-(r/r\_blend))
                \State d2\_eval[j][0] = v
            \EndIf
        \EndFor
    \EndFor
    \State Communicate()
\end{algorithmic}
\end{algorithm}

\begin{algorithm}
\caption{Build buffer region}
\label{alg:build_buffer_region}
\begin{algorithmic}[1]
\For{i $\leftarrow$ blended atoms}
    \For{j $\leftarrow$ neighbors of atom i}
        \If{$||\mathrm{pos(i)} - \mathrm{pos(j)}||$ $<$ r\_buffer}
            \State i2\_potential[j][0] = 1
        \EndIf
    \EndFor
\EndFor
\State Communicate()
\end{algorithmic} 
\end{algorithm}

\subsection{Spatial potential mixing}\label{sec:force_mixing}
The mixing scheme implemented in ML-MIX is force-mixing, which is a scheme commonly used in QM/MM. For in-depth reviews of force-mixing in the QM/MM context, please see Bernstein et al \cite{Bernstein_2009}. Here we briefly review the force-mixing method, and discuss simplifications that can be made when using two local potentials. \\

% \subsubsection{Force mixing}

\begin{figure}
    \centering
    \includegraphics[width=\linewidth]{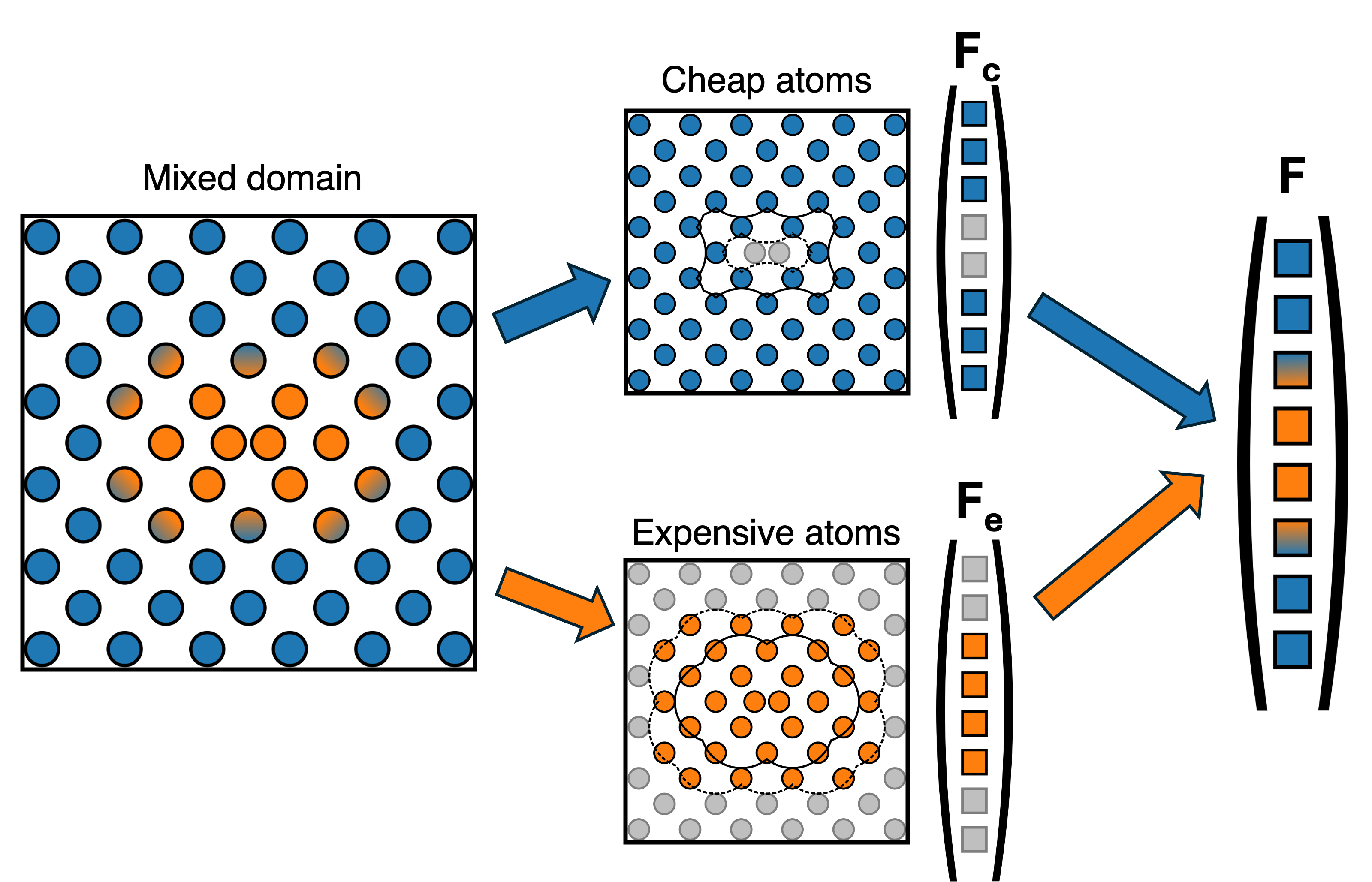}
    \caption{An illustration of force-mixing with interatomic potentials. A mixed domain is split into two non-interacting subdomains \TC{(blue and orange)}, which are composed of all the atoms in the core and blending regions for each potential (solid line), plus an extra buffer region containing atoms up to the potential cutoff radius out (dotted line). Partial force vectors $\mathbf{F}_{\mathrm{c}}$ and $\mathbf{F}_{\mathrm{e}}$ are obtained by evaluating the cheap and expensive subdomains respectively. These partial vectors are then combined into the full force vector on the system $\mathbf{F}$. In this process, forces on buffer atoms are discarded and forces on blending atoms are mixed between potentials.}
    \label{fig:force_mixing}
\end{figure}

The process of force mixing is schematically shown in Fig.~\ref{fig:force_mixing}. To obtain the forces, the domain is split into separate segments where an additional buffer region is included to stop individual regions seeing an artificial surface. The forces on these regions are then evaluated to create separate parts of the overall force vector. The full force vector is constructed as a combination of these separate vectors. In QM/MM, as DFT is a non-local method, the size of the buffer region around the QM domain is a crucial convergence parameter - one needs to balance the error due to the separation of the region of interest from the artificial surface against additional cost incurred as atoms are added. \\

Local potentials simplify this. Consider a set of atoms $\Lambda_{k}$, which is a subset within a larger domain $\Lambda_{k} \subseteq \Lambda$. We want to attain correct forces on atoms in $\Lambda_{k}$ with a local potential $\Phi_{k}$ that has a defined cutoff radius $r_{\mathrm{cutoff}}$. The force on atom $i \in \Lambda_{k}$ can be written in terms of local energies as
\begin{equation}
    F_{i} = \frac{\partial E}{\partial \mathbf{x}_{i}} = \sum_{j}^{j\in \mathrm{neigh}(i)}{\frac{\partial E^{k}_{j}}{\partial \mathbf{x}_{i}}}
\end{equation}
where $E^{k}_{j}$ represents the local energy of atom $j$ as evaluated by $\Phi_{k}$. Due to the local nature of the potential, we only need to evaluate $E^{k}_{j}$ for atoms that lie within $r_{\mathrm{cutoff}}$ of atom $i$. For an atom that lies on the edge of $\Lambda_{k}$, it is clear that attaining the correct force requires evaluation of atoms up to $r_{\mathrm{cutoff}}$ outside $\Lambda_{k}$. Defining the distance between two atoms $i, j$ as $d(i, j)$, the buffer region $\Lambda_{k\text{-buffer}}$ is given by

\begin{equation}
\begin{split}
\Lambda_{k\text{-buffer}} = \Big\{ i \in \Lambda \setminus \Lambda_k \,\Big|\,  
\exists j \in \Lambda_k \\
\text{ and } d(i, j) \leq r_{\mathrm{cutoff}} \Big\}.
\end{split}
\end{equation}

Note that through the dependence of local energies on local atomic positions, there is an implicit dependence on atomic positions up to two cutoff radii out. Hence, \texttt{r\_buffer} need not be any greater than the potential \rebuttal{receptive field, as at this radius the forces match what they would be in non-mixed simulations on each domain. For MLIP architectures with a large receptive field due to message passing (e.g, MACE), one often does not need a buffer that captures it in its entirety; force errors reduce in size rapidly with buffer radius.}\\

Force mixing is implemented in ML-MIX through a wrapper \texttt{pair\_style} which evaluates multiple sub-\texttt{pair\_styles} in turn. Algorithm \ref{alg:force_mixing} demonstrates this process. \\

\begin{algorithm}
\caption{Force mixing}
\label{alg:force_mixing}
\begin{algorithmic}[1]
\State Initialise forces to 0
\For{i $\leftarrow$ potentials}
    \For{j $\leftarrow$ atoms}
        \If{i2\_potential[i][j] == 0}
            \State Remove j from neighbor list
        \EndIf
    \EndFor
    \State temp\_forces = potentials(i) $\rightarrow$ compute()
    \State temp\_forces = temp\_forces * d2\_eval[i][:]
    \State forces = forces + temp\_forces
    \State Restore neighbor list
\EndFor
\end{algorithmic} 
\end{algorithm}

\subsection{Si vacancy migration}
To generate the vacancy structure, a 10$\times$10$\times$10 (8000) atom Si supercell was generated, and the atom nearest the centre was removed. Initial guesses for the minimum energy pathway were formed by linearly interpolating the migration of a neighboring atom into this vacancy. Including the start and end states, 9 images were generated along this pathway, which were then relaxed using the nudged-elastic-band (NEB) method \cite{Henkelman_2000} as implemented in \texttt{LAMMPS}. Further analysis was conducted in Python using ASE \cite{ASE}. For the ML/ML simulation, the seed atoms for the expensive potential region were a union of all atoms immediately surrounding the vacancy in the final and end states. These neighbors were identified using coordination analysis with a cutoff of 4.0~{\AA}, with all atoms that have 15 neighbors being selected. An additional 4.0~{\AA} of core region was added around these atoms, as well as 4.0~{\AA} of blending region. 6.0~{\AA} of buffer region was then included, which matches the cutoff of the cheap and expensive potentials. Region rebuilding was switched off\rebuttal{; provided expensive atoms do not move far this allows relaxations to progress smoother and faster by stopping atoms changing the proportion of forces they gain from each potential as the minimization progresses}. For both the all-expensive and ML/ML NEB simulations, no energy tolerance was set (as total energies are not present in the ML/ML simulation) and the force tolerance was set such that $\|\mathbf{F}\|_{\infty} < 0.0005$~eV/{\AA}. Tighter tolerances than this would frequently fail to converge for the ML/ML simulation. If tighter tolerances are necessary, then we recommend switching to the all-expensive regime for the last steps of a relaxation. The final energies were obtained through one-shot energy evaluations of the relaxed structures with the expensive potential.

\subsection{Si stretched bond}
To generate the structure, a $10\times 10 \times 10$ (8000 atom) periodic supercell of bulk Si was constructed and a single bond was stretched an additional 0.1~{\AA} along its length and fixed using the rigid package in \texttt{LAMMPS} \cite{Kamberaj_2005}. This block was then pre-thermalized to 300 K in a 2 ps simulation with a Langevin thermostat using only the expensive potential. This thermalized initial structure was used as the starting point for all subsequent comparison simulations, in which the average force on this bond was recorded over 100 ps. For the all-expensive simulations, an $NVE$ ensemble was used. For the ML/ML simulations, the seed atoms were selected to be the atoms involved in the stretched bond, no blending region was used (abrupt force mixing), as well as a weak thermostat of time constant 2.0 ps applied to only the cheap potential region in order to stop temperature drift due to the ML/ML boundary . For all simulations a buffer region size of 6.0~{\AA} was used (which matches the cutoff of both potentials). The size of the core region was varied between simulations, as described in section.~\ref{sec:si_stretched_bond}. The expensive potential region was rebuilt about this seed atom every timestep. \rebuttal{To compute the average bond force, samples of atomic forces were taken every 15~fs, long enough to avoid significant correlation between samples.} \\

\subsection{Diffusion coefficients}
\label{sec:diffusion}
To calculate the diffusion coefficients, individual defects were placed into periodic cells of size 16 $\times$ 16 $\times$ 16 unit cells ($\sim 8000$ atoms). Cells were heated to the target simulation temperature over 2~ps with a Langevin thermostat. For the Fe simulation, diffusion coefficients were measured at 800~K, 900~K, 1000~K and 1100~K. For W--He, diffusion coefficients were measured at 400~K, 600~K and 800~K. The reference simulations were then switched to NVE, whilst the ML/ML \rebuttal{simulations had a weak Langevin thermostat (damping parameter - 2.0 ps) applied only to the cheap potential region}. For the Fe simulation, each data point consisted of the average of five diffusion coefficients derived from the mean square displacements from non-overlapping time-lags of $5\times1$~ns simulations. For the W--He simulations, each point was the average of twenty-five diffusion coefficients derived from the mean square displacements from non-overlapping time-lags from $25\times60$~ps simulations. In all cases, the error was estimated from the standard error in the averaged diffusion coefficients. For the W--He ML/ML simulation, the seed atom was selected to be the single He present in the simulation, and regions were rebuilt around this seed atom every timestep. For the Fe ML/ML simulation, seed atoms were selected by finding the most highly coordinated atoms (on average) in the simulation every 100 timesteps. For this, measurements of atomic coordination $Z$ (with a cutoff of 2.0~{\AA}) were taken every 10 timesteps, and these were averaged using \texttt{fix ave/atom}. New seed atoms were selected according to the criteria that $\langle Z \rangle >0.4$. Regions were rebuilt around these seed atoms every timestep. For the first 100 timesteps in the Fe ML/ML simulation, only the expensive potential was used. Both simulations used a core potential region of 6.0~{\AA} around the defect atoms, as well as a blending region of 4.0~{\AA}. For the Fe potential, a buffer width of 5.5~{\AA} was used, whilst for the W--He potential, a width of 6.0~{\AA} was used, matching the UF3 potential cutoff.

\rebuttal{\AWE{\subsection{W dislocation glide}}}
\rebuttal{\AWE{Simulating the thermally activated glide of W screw dislocations is known to require $\approx10^5$ atoms due to finite-size effects \cite{cereceda_2012,allera_2024, Bacon_2009}. These are (i) in $x$, the cell length must be large enough to prevent self-interaction and to ensure that the work done can dissipate without substantial heating, (ii) in $y$, the surface must be far enough away to prevent attraction and dislocation annihilation, and (iii) in $z$, to ensure that the dislocation moves by the correct mechanism of the nucleation and propagation of kink pairs. In this work, the dimensions of the simulation cell in the glide direction (along $ x  = [10\overline{1}]$), glide plane normal (along $ y  = [\overline{1}2\overline{1}]$), and dislocation line (along $ z  = [111]$) were chosen to be $223~\mathrm{\AA} \times 221~\mathrm{\AA} \times 68~\mathrm{\AA}\ (25~\textit{b}$), respectively, for a total of 195,000 atoms. At each temperature simulated, the thermally expanded lattice parameter was measured and the cell was rescaled to match. In each simulation, the dislocation was placed at the centre of the simulation cell, where the atomic displacement field was generated using the matscipy dislocation module \cite{matscipy}. Periodicity was imposed in $x$ and $z$, with free surfaces in $y$. Shear stresses $\sigma_{\mathrm{yz}}$ were applied to these free surfaces, simulating \{112\} loading. During glide, the dislocation moved on alternating \{110\} planes. \\}}

\rebuttal{\AWE{Using a 2~fs timestep, the cell was equilibrated to the chosen temperature for 50~ps with a Langevin thermostat. Following this, a 250~ps production run was conducted, from which dislocation positions were extracted (using OVITO dislocation analysis \cite{Stukowski_2010}) and an average dislocation velocity was calculated. In the all-expensive simulations, an NVE ensemble was used, whilst in the mixed simulations, a weak Nosé-Hoover thermostat was applied to the whole system (damping time 10 ps) in order to counteract the gradual heating that occurred due to energy flux from the ML/ML boundary. A short initial test was conducted to ensure that this thermostat did not impact dynamics. \\}}

\rebuttal{\AWE{The dislocation core was identified as the expensive region using Common Neighbor Analysis (CNA) in LAMMPS. The radius of the core, blending and buffer regions were 5.0, 4.0 and 5.0 $\mathrm{\AA}$, respectively. The expensive region corresponded to $\approx1\%$ of the total atoms in the simulation cell.\\}}

\rebuttal{\subsection{He implantation into W}} \label{sec:he_implantation}

\begin{table}[]
\caption{The cell size and number of atoms used for He implantation simulations at energies up to 100~eV.}
\label{tab:implantation_cell_sizes}
\begin{tabular}{l|ll}
He Energy (eV) & Cell Size ({\AA}) & Number of Atoms \\ \hline
5-30 & 39.8$\times$39.8$\times$46.1 & 5070 \\
35-40 & 46.1$\times$46.1$\times$62.0 & 9000 \\
50-60 & 46.1$\times$46.1$\times$93.8 & 13500 \\
70 & 46.1$\times$46.1$\times$125.6 & 18000 \\
80 & 46.1$\times$46.1$\times$157.4 & 22500 \\
90 & 46.1$\times$46.1$\times$189.2 & 27000 \\
100 & 46.1$\times$46.1$\times$221.0 & 31500
\end{tabular}
\end{table}
\rebuttal{For the study of He implantation into W, a modified version of the procedure outlined in Borovikov et al \cite{Borovikov_2014} was followed. We targeted normal deposition into 1000~K (001) W surfaces. A slab-type simulation domain was selected, which was non-periodic in $z$ [001] and periodic in $x$ [100] and $y$ [010]. First, preliminary studies were run to determine the necessary size of the slab. It was found that high energy He atoms ($> 40$~eV) can penetrate very deeply into W samples, in some cases up to 4$\times$ the mean depth. It was also found that cells needed to be made wide enough that a deflected, obliquely penetrating He atom that passed through periodic boundaries did not pass close to atoms it had previously disturbed. Table~\ref{tab:implantation_cell_sizes} contains the cell sizes and number of atoms used for each He implantation energy. All simulations were carried out in LAMMPS, with the following procedure. Firstly, the slabs were thermalized to 1000~K over 10~ps of MD, with a 1~fs timestep using a Langevin thermostat with time constant of 0.01~ps. For computational efficiency, the same initial thermalized slab configuration was used for sets of implantations carried out in parallel. In all following simulations (both all-expensive and mixed), the thermostat was removed and the molecular dynamics was run using an NVE ensemble.\\}

\rebuttal{To generate the initial configurations for each implantation, a He atom was placed in a uniformly sampled random position 5.0~{\AA} above the W surface and given a velocity in $- z$ corresponding to the investigated kinetic energy (between 5 and 100~eV). An initial simulation timestep was selected corresponding to an initial He per-timestep displacement of 0.02~{\AA}. The initial surface impact simulation was then run for 750 timesteps, which would correspond to a total straight-line He distance of 15.0 {\AA}. At the end of this, the He atom position was measured. If it was found to be more than 2.0 {\AA} above the surface, and with a positive $z$ velocity, the simulation was stopped with the result being that the atom reflected. If not, the timestep was updated to either 0.02 {\AA} divided by average He velocity or 0.1 fs, whichever was smaller, and a further 0.05~ps was run. This was repeated until either the He escaped the surface or a total time of 1~ps passed. For all energies, 1~ps was enough time for the He atom to thermalise.\\}

\rebuttal{All expensive simulations were carried out with He energies up to 40 eV. For each energy, 4 sets of 25 implantation attempts were carried out, with each set of 25 sharing the same thermalized slab structure.\\}

\rebuttal{For the mixed simulations, the expensive region was set to track the He atom, with a core radius of 6.0 {\AA} a blending radius of 4.0 {\AA} and a buffer radius of 8.0 {\AA}. The thermalization was entirely conducted using the cheap potential. For all energies, 5 sets of 100 repeats were used.\\}

\rebuttal{The potentials were evaluated in LAMMPS on 40 GB A100 GPUs using the Kokkos \cite{kokkos1, kokkos2} accelerated variants of the \texttt{symmetrix} package for MACE \cite{Kovacs_2025_maceoff} and \texttt{ML\_SNAP} package for SNAP \cite{Thompson_2015}. In order to conduct mixed simulations where both the expensive and cheap potentials were evaluated on the same GPU, a Kokkos version of ML-MIX was developed. This can be found in v0.3 of ML-MIX onwards, and allows for force-mixing on GPU of many kokkos-compatible \texttt{LAMMPS} pair\_styles.}

\section*{Data availability}
Data generated and analysed in the current study is available in the \rebuttal{v0.3} version of the ML-MIX repository released on Zenodo \cite{ml-mix-zenodo}.

\section*{Code availability}
The ML-MIX package along with scripts necessary to reproduce the results presented here are available in the ML-MIX GitHub repository \cite{ML-MIX}.

\section*{Acknowledgements}
We thank Christoph Ortner \rebuttal{and Albert P. Bartók} for helpful discussions. We also thank Lakshmi Shenoy for the Fe fitting script. FB \rebuttal{and MN} are supported by studentships funded by the UK Engineering and Physical Sciences Research Council supported Centre for Doctoral Training in Modelling of Heterogeneous Systems, Grant No. EP/S022848/1. \rebuttal{MN is also supported by AWE Nuclear Security Techonologies.} We acknowledge the University of Warwick Scientific Computing Research Technology Platform for assisting the research described within this study. Some of the calculations were performed using the Sulis Tier 2 HPC platform hosted by the Scientific Computing Research Technology Platform at the University of Warwick. Sulis is funded by EPSRC Grant EP/T022108/1 and the HPC Midlands+ consortium. \rebuttal{We are grateful for computational support from the UK national high performance computing service, ARCHER2, for which access was obtained via the UKCP consortium and funded by EPSRC Grant No. EP/X035891/1.} TDS gratefully acknowledges support from ANR grants ANR-19-CE46-0006-1, ANR-23-CE46-0006-1, IDRIS allocation A0120913455 and an Emergence@INP grant from the CNRS. 

\section*{Competing interests}
All authors declare no financial or non-financial competing interests. 

\section*{Contribution statement}
FB, \rebuttal{MN,} TDS and JRK designed the research. FB developed the ML-MIX software, generated the data and performed the analyses \rebuttal{for all examples except the W screw dislocation glide, which was conducted by MN}. TDS and JRK provided supervision, guidance, and feedback throughout. FB wrote the paper, with input from TDS and JRK at all stages, \rebuttal{as well as MN at the final stage}. \rebuttal{A draft of the W screw dislocation sections was provided by MN.} All authors revised the paper and approved the final version.

% \bibliography{sn-bibliography}
\printbibliography
\end{document}

\title[Article Title]{Supplementary Information for `Efficient and Accurate Spatial Mixing of Machine Learned Interatomic Potentials for Materials Science'}

\author*[1]{\fnm{Fraser} \sur{Birks}}\email{fraser.birks@warwick.ac.uk}

\author[1,2]{\fnm{Matthew} \sur{Nutter}} \email{}

\author[3]{\fnm{Thomas} D \sur{Swinburne}}\email{}

\author[1]{\fnm{James} R \sur{Kermode}}

\affil*[1]{\orgdiv{Warwick Centre for Predictive Modelling, School of Engineering}, \orgname{University of Warwick}, \orgaddress{\street{Library Road}, \city{Coventry}, \postcode{CV4 7AL}, \country{United Kingdom}}}

\affil[2]{\orgdiv{Department of Physics}, \orgname{University of Warwick}, \orgaddress{\street{Library Road}, \city{Coventry}, \postcode{CV4 7AL}, \country{United Kingdom}}}

\affil[3]{\orgdiv{Aix-Marseille Université}, \orgname{CNRS}, \orgaddress{\street{CINaM UMR 7325, Campus de Luminy}, \city{Marseille}, \postcode{13288}, \country{France}}}

\maketitle

\section{Linear ACE potential fitting parameters}\label{secS1}
Here we present the full set of potential fitting parameters for the Fe, W and W-He ACE potentials used in the study. The main fitting parameters are presented in Table~\ref{tab:potential_params}. Weights by config type for the Fe potential are shown in Table~\ref{tab:Fe_ACE_weights}. For the W and W--He ACE, we derive energy weights for configurations following the method suggested in \cite{Witt_2023} in which the energy weight on configuration $R$ is generated by 
\begin{equation}
    w_{R{}} = \frac{1000}{\sqrt{N_{R}}}
\end{equation}
where $N_{R}$ is the number of atoms in configuration $R$. For each configuration, the force weights were set to 10.0 and the virial weights were set to 0.0. Virials were not used in the fitting in order to avoid any errors introduced from underconverged (insufficient $K$-point density) DFT, as suggested by the dataset author \cite{Nutter_2024_supp}. The exception to this were the single W atom configurations (which were evaluated at a tighter $K$-point density), which had an energy weight of 4000.0, a force weight of 100.0 and a virial weight of 200.0.

\begin{table}[]
\caption{Fe, W and W--He ACE potential parameters. BLR stands for Bayesian linear regression.}
\begin{tabular}{l|lll}
 & Fe ACE & W ACE & W--He ACE \\ \hline
Cutoff & 5.5 & 5.0 & 5.0 \\ 
Corr. order & 3 & 3 & 3 \\ 
Max. poly. deg. & 20 & 21 & 20 \\
Solver & BLR & BLR & BLR \\ 
\begin{tabular}[c]{@{}l@{}}Smoothness prior\\ strength\end{tabular} & 4 & 4 & 4
\end{tabular}
\label{tab:potential_params}
\end{table}

\begin{table}[]
\caption{Fe ACE configuration weighting.}
\begin{tabular}{l|lll}
Config type & \begin{tabular}[c]{@{}l@{}}Energy\\ weight\end{tabular} & \begin{tabular}[c]{@{}l@{}}Force\\ weight\end{tabular} & \begin{tabular}[c]{@{}l@{}}Virial\\ weight\end{tabular} \\ \hline
default & 30.0 & 1.0 & 1.0 \\ 
slice\_sample & 80.0 & 0.1 & 1.0 \\ 
prim\_random & 80.0 & 0.1 & 1.0 \\ 
\end{tabular}
\label{tab:Fe_ACE_weights}
\end{table}

\section{Speedup and load balancing of mixed simulations}\label{secB1}
This section provides an investigation into the speedups for load balanced and non-load balanced ML-MIX simulations in silicon with a fixed expensive potential region size for (a) a fixed number of atoms and an increasing number of processors and (b) a fixed number of processors and an increasing number of atoms. It is shown that load balancing is most important when considering small domains ($<$30,000 atoms) and fewer MPI parallel processors ($<$15). It is shown that for small domains, the speedup variability between different numbers of processors is lessened by load balancing. \\

% \subsection{Method}
MD simulations were conducted with Si domains at 0 K (frozen atoms). Two sets of simulations were run, set (a) where the number of atoms were fixed and the number of processors were increased between 1 and 48, and set (b) where the number of processors was fixed and the number of atoms were increased between $8 \times 10^{3}$ and $10^{6}$. For each separate simulation domain, timing measurements were carried out with (i) only the 2\_10 ACE Si cheap potential, (ii) only the 4\_20 ACE Si expensive potential, (iii) load-balanced and (iv) non-load balanced ML/ML simulations containing both the expensive and cheap potentials. Within each simulation domain, the same number of timesteps was used for each timing measurement (i-iv), with the number selected such that the wall-time of the longest (the all-expensive) simulation was approximately 45 minutes. To build the expensive domain in each ML/ML simulation, single seed atoms were selected at the center of each cell, around which 6~{\AA} core regions and 6~{\AA} buffer regions were constructed. This meant that the number of expensive potential atoms was fixed across all ML/ML simulations, regardless of overall domain size. \\

For set (a), two atomistic domain sizes were chosen; a `small' domain of 8000 atoms, and a `large' domain of 262,144 atoms. For set (b), the number of processors was set at 27 for all domain sizes used. For each simulation domain investigated, the `upper-limit' speedup $S_{\mathrm{UL}}$ was computed with 
\begin{equation} \label{equation:speedup_theo}
    S_{\mathrm{UL}} = \frac{N S_{\mathrm{C/E}}}{N_{\mathrm{E}} S_{\mathrm{C/E}} + N_{\mathrm{C}}},
\end{equation}
where $N$ is the total number of atoms in a simulation domain, $N_{\mathrm{E}}$ and $N_{\mathrm{C}}$ are the number of expensive and cheap atoms (including those in both buffer regions), and $S_{\mathrm{C/E}}$ is the measured speedup of the all-cheap simulation over the all-expensive simulation in the domain of interest. \\

Simulations were carried out on Dell PowerEdge C6420 compute nodes each with 2 $x$ Intel Xeon Platinum 826 (Cascade Lake) 2.9 GHz 24-core processors; 48 cores per node; 192 GB DDR4-2933 RAM per node; 4 GB RAM per core.

\subsection{Load balancing strategies}
Load-balancing is performed via the \texttt{fix balance} command in \texttt{LAMMPS}. If the \texttt{time} keyword is specified \texttt{LAMMPS} attempts to dynamically decompose the overall domain into subdomains of equal computational cost. \texttt{LAMMPS} has two strategies to decompose domains - `brick' and `tiled'. With the `brick' strategy, each individual sub-domain is cuboidal, and is constrained to be joined at the corners. With the `tiled' strategy, each domain is also cuboidal, but not necessarily constrained to connect at corners. For more details, please refer to the \texttt{comm\_style} page of the \texttt{LAMMPS} documentation \cite{comm_style}. Within the \texttt{fix balance} command, two algorithms are available; the \texttt{shift} algorithm, compatible with both the `brick' and `tiled' decomposition strategies, and the \texttt{rcb} algorithm, compatible with just the `tiled' decomposition strategy. For more detail on these algorithms, please refer to the \texttt{fix balance} page of the \texttt{LAMMPS} documentation \cite{fix_balance}. Each load balancing strategy was investigated, and it was found for a simple domain with one central expensive region, the \texttt{shift} algorithm performed marginally better when domains were small and the \texttt{rcb} algorithm performed marginally better when domains were large. Throughout this paper, the \texttt{shift} algorithm is used with the brick domain decomposition strategy. We recommend users test each load-balancing strategy to determine which works best for their system.

\subsection{Results and discussion}
\subsubsection{Fixed atoms, increasing processors}
\begin{figure*}
    \centering
    \includegraphics[width=0.9\linewidth]{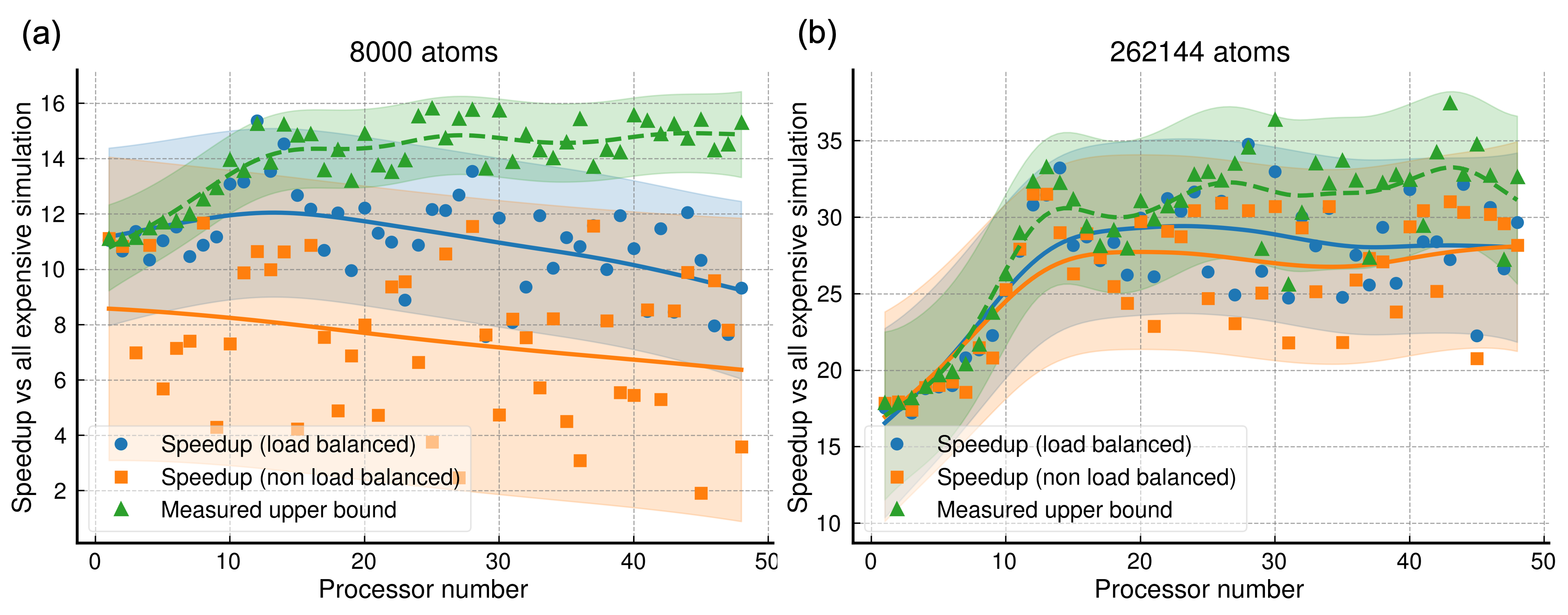}
    \caption{Speedup of ML/ML simulations over all-expensive reference simulations on an increasing number of processors for two domain sizes, \TC{both with load-balancing (blue points) and without (orange points)}. In each case, the upper bound speedup (green line) is determined from equation \ref{equation:speedup_theo}. \TC{(a)}: Speedup on a small (8000 atom) domain. For this domain size, one can only attain near-upper bound speedups for small numbers of processors if a load-balancing strategy is used. The effect of load balancing is very pronounced, as without it all expensive atoms frequently end up on one sub-domain, creating simulation bottle-necks. \TC{(b)}: Speedup on a large (262,144 atom) domain. At this domain size, one attains a near upper-bound speedup for any number of MPI parallel domains, and load balancing has a relatively minor effect. All trend-lines are fitted with Gaussian process regression.}
    \label{fig:fixed_atom_load_balancing}
\end{figure*}
The results for simulation set (a) are displayed in Fig.~\ref{fig:fixed_atom_load_balancing}. For the smaller (8000 atom) domain, it can be seen that with load balancing, a near-upper-bound speedup can be attained for a number of processors below $\sim 15$. Above this, the load balanced simulation falls significantly below the upper-bound due to communication overheads taking up a significant fraction of the runtime. Without load balancing, the performance is on average significantly worse, but varies greatly between the number of processors chosen. Particularly poor performances are due to all of the expensive atoms landing on a single parallel domain. For the larger (262,144 atom) domain, it can be seen that near-upper bound speedup is approximately attained for any processor number, regardless of load-balancing. This arises due to every individual subdomain always being far larger than the expensive potential region. 

\subsubsection{Fixed processors, increasing atoms}
\begin{figure}
    \centering
    \includegraphics[width=0.6\linewidth]{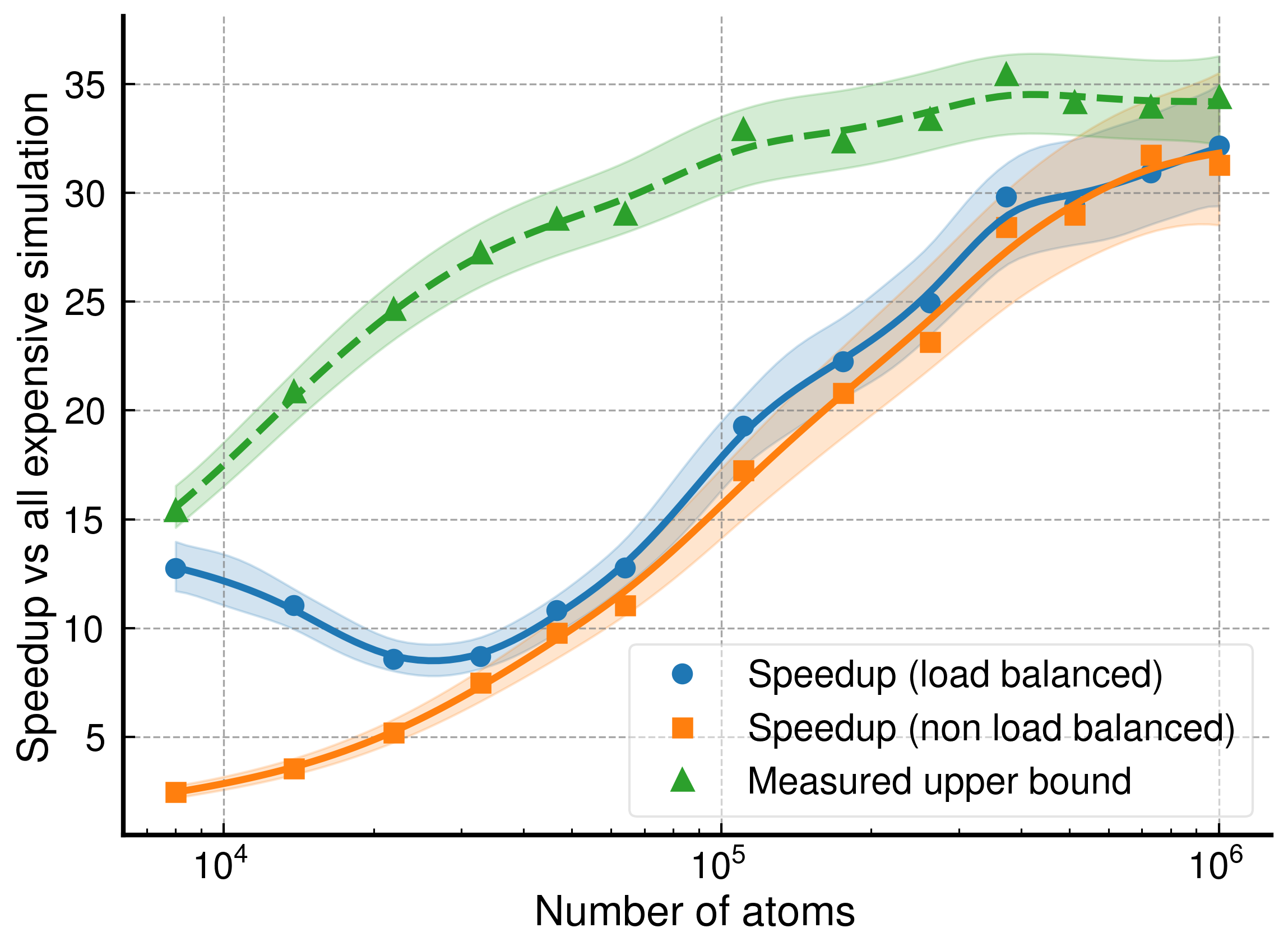}
    \caption{The measured speedup of the mixed simulation over the all expensive simulation on 27 processors, for increasing numbers of atoms \TC{both with load-balancing (blue points) and without (orange points)}. The green line indicates the maximum possible speedup (with no overheads), computed from equation \ref{equation:speedup_theo}. It can be seen that for small numbers of atoms ($<$30,000) load balancing is important, and for large number of atoms ($\sim 10^{6}$) the speedup tends toward the upper bound. Trend lines have been fitted with Gaussian process regression.}
    \label{fig:load_balance_fixed_proc}
\end{figure}
Results for simulation set (b) in which the simulation domain size is increased at a fixed number of processors (27) is shown in Fig.~\ref{fig:load_balance_fixed_proc}. For smaller atomic domains ($<$30,000 atoms), it can be seen that load balancing is important, providing a considerable speedup. As the size of the domain gets larger, load-balancing becomes less important, and the overall speedup tends toward the overhead-free upper limit.

\section{Impact of constrained fitting}\label{secC1}
In this paper, the cheap potentials have been constrained to enforce elastic constant matching. In this section, we show that matching the elastic constants between our potentials leads to reduced errors in ML-MIX relaxed structures for less than 3\% strain deviations from the bulk. \\

% \subsection{Method}
A periodic $10 \times 10 \times 10$ (8000 atom) bulk Si structure was constructed and sheared in the $xy$ to directions to strains between 0 and 7\%. For each structure, a joint ML/ML relaxation was performed using ML-MIX. The central atom and the surrounding 6~{\AA} were modeled expensively, with the remainder of the atoms were with the constrained or unconstrained Si 2\_10 cheap potentials in two separate investigations. For the relaxation, a blending region of 6.0~{\AA} was used, as well as a relatively loose tolerance of $\|\mathbf{F}\|_2 < 10^{-3}$~eV/{\AA}. Smaller tolerances than this would frequently fail to converge. The strain error $Du$ was then found for each structure as
\begin{equation}
    Du = \|\nabla u - \nabla u_{\mathrm{ref}}\|_2
\end{equation}
where $u$ is the displacement with respect to the ML/ML relaxed structure and $u_{\mathrm{ref}}$ is the displacement field from the relaxed all-expensive reference.

\subsection{Results and discussion}
\begin{figure}
    \centering
    \includegraphics[width=0.6\linewidth]{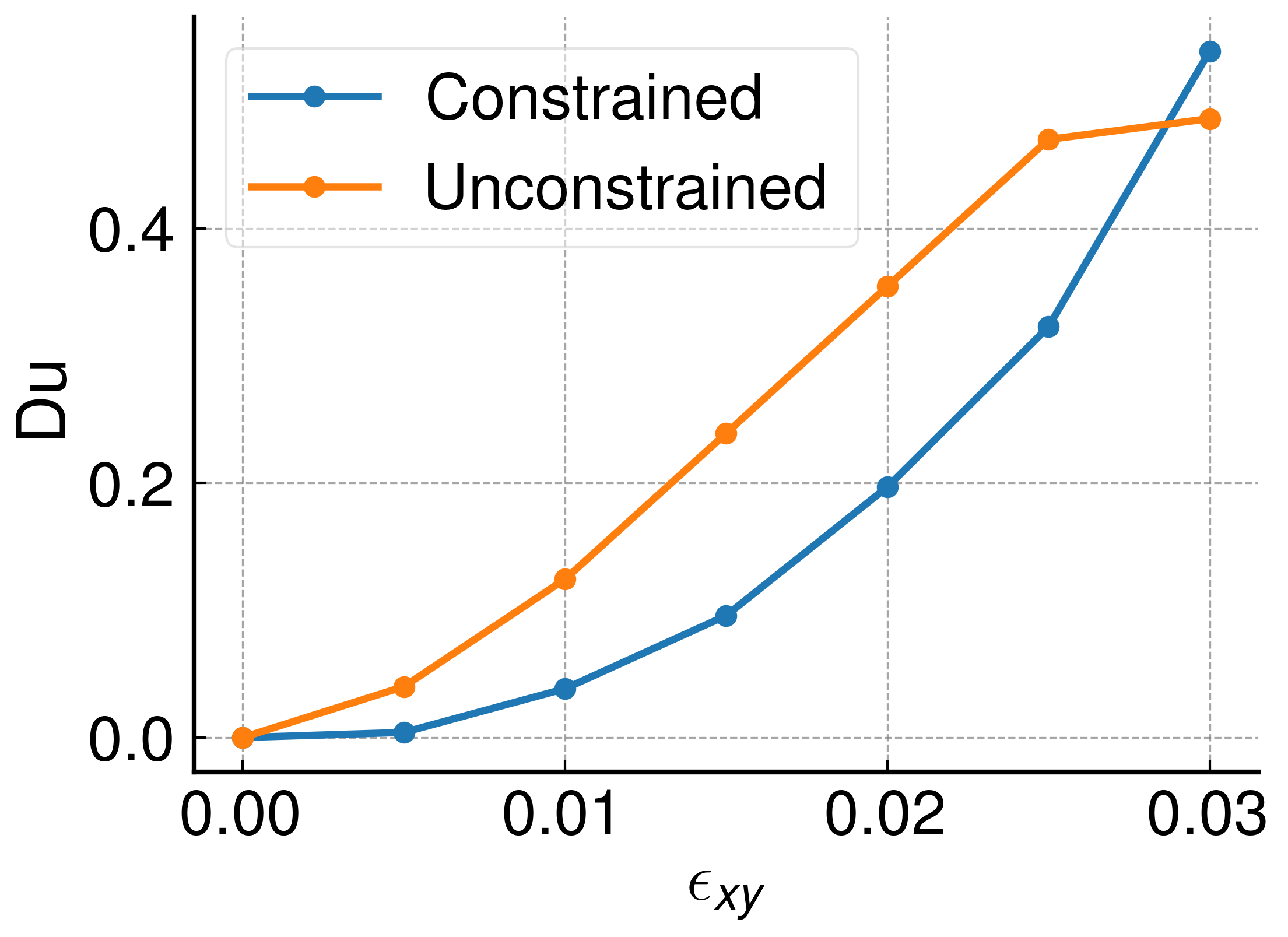}
    \caption{The strain error norm Du measured for sheared and relaxed bulk Si \TC{ML/ML} simulations using \TC{cheap potentials which are a both constrained (blue points) and unconstrained (orange points)} for different applied shear strain. \TC{The constrained cheap potential has elastic constants that closely match the expensive potential.} The ML/ML simulation contains a central 6~{\AA} sphere of the expensive potential surrounded by the cheap potential. $Du$ is far greater for small strains when using the unconstrained cheap potential, as the mismatched elastic constants to the expensive lead to Eshelby-inclusion type strain fields.}
    \label{fig:constrained_potentials}
\end{figure}

Fig.~\ref{fig:constrained_potentials} shows the strain error in the ML/ML relaxed structures. The spatial distribution of the error in the strain fields matches that expected from an Eshelby inclusion, and it can be seen that for very small strains $\epsilon_{xy}<0.005$, the constrained potential exhibits very little error in the strain whilst the constrained potential error increases immediately. The constrained potential continues to have lower strain error up until $\epsilon_{xy}=0.03$.

\section{Energy conservation}\label{secD1}
A fundamental limitation of force mixing is a lack of energy conservation. Whilst this has been noted previously \cite{Bernstein_2009}, the features in a force-mixed simulation that may worsen or improve energy conservation are unclear. Here we present a brief investigation where we test the impact on energy conservation of (i) having a larger interface region between potentials, (ii) having a more flexible cheap potential and (iii) introducing a larger blending region between potentials. Whilst findings presented here are derived from ML/ML simulations using ML-MIX, the findings are applicable to force-mixing simulations more generally, including QM/MM simulations. \\

% \subsection{Method}
A periodic cube of Si was constructed which matches that described in section 2.3 of the main manuscript. The spherical central region of the domain (around the fixed stretched bond) was simulated expensively, with the remainder of the domain modeled cheaply. The expensive potential was the same Si ACE used in section 2.3 of the main manuscript. Three different cheap ACE potentials were used, each fit according to the constrained fitting procedure outlined in the main manuscript. The key parameters of these potentials are displayed in Table~\ref{tab:eng_consv_pot_params}. To investigate energy drift in the simulation set-up, a 50~ps, 300~K ML-MIX \texttt{LAMMPS} MD simulation was performed with a timestep of 1 fs. As the total energy of a force-mixed simulation is not defined, the total energy of the full system was measured through evaluating snapshots output every ps with the expensive potential. \\

\begin{table}[]
\caption{Si cheap potential parameters.}
\begin{tabular}{l|lll}
 & 2\_10 & 2\_15 & 3\_15 \\ \hline
Cutoff & 6.0 & 6.0 & 6.0 \\
Corr. order & 2 & 2 & 3 \\
Max. poly. deg. & 10 & 15 & 15 \\ 
\end{tabular}
\label{tab:eng_consv_pot_params}
\end{table}

Two separate investigations were conducted for each cheap potential. (a) For a set-up with no blending region (abrupt force-mixing) the size of the spherical expensive potential region was varied between 4 and 10~\AA, and (b) for a fixed 10~{\AA} core + blending region, the amount of blending region is varied between 4 and 6~{\AA}. Both linear and cubic blending were investigated. \\

\subsection{Results and discussion}
\subsubsection{Size of potential-potential interface}
\begin{figure*}
    \centering
    \includegraphics[width=0.9\linewidth]{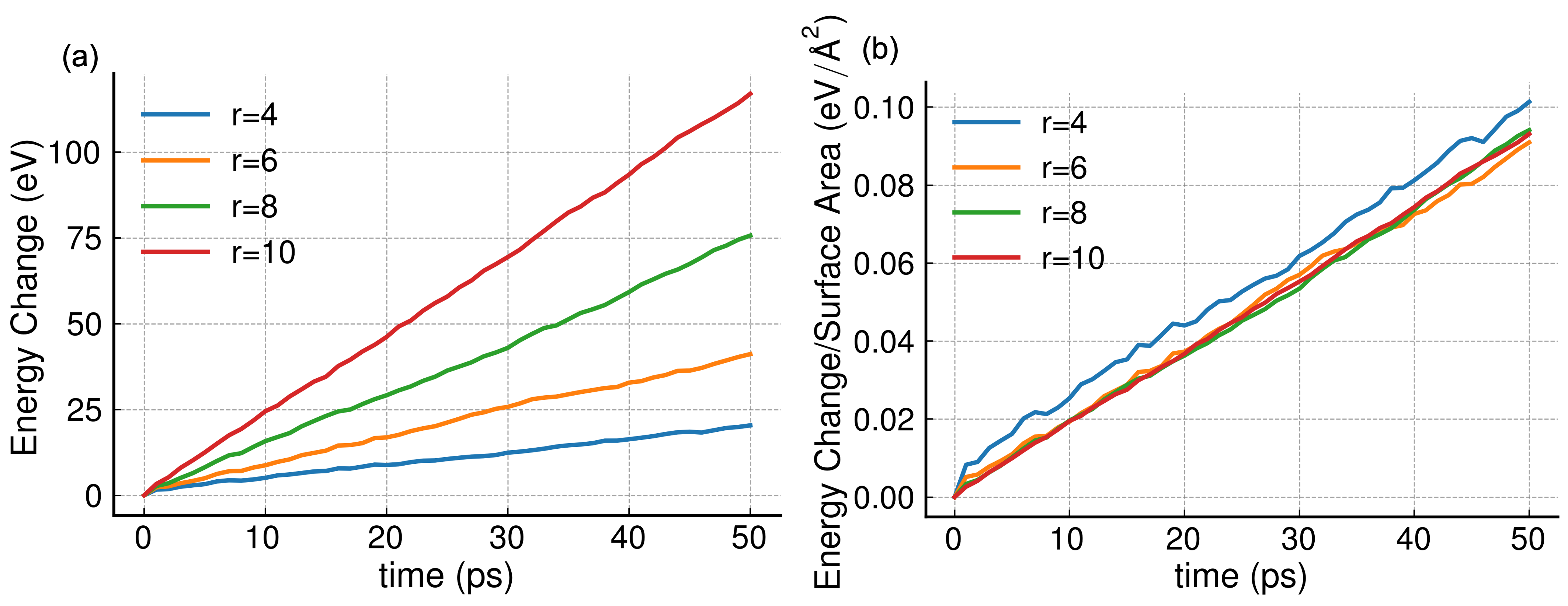}
    \caption{Energy drift over time for an abruptly mixed ML/ML simulation of bulk Si, using the 2\_10 cheap potential. The energies are that of the full system, computed at snapshots every ps with the expensive potential. \TC{Blue, orange, green and red curves refer to sizes of the expensive potential region of 4, 6, 8 and 10~{\AA} respectively.} \TC{(a)}: The raw energy change for each simulation, showing that a larger expensive potential region causes a greater rate of energy increase. \TC{(b)}: Energy change normalized by spherical interfacial contact area between potentials, showing that for a given two potentials in abrupt contact the rate of energy change per unit of contact area (energy flux) is constant.}
    \label{fig:energy_drift}
\end{figure*}

The energy over time for the abruptly force-mixed ML/ML simulations are shown on the left of Fig.~\ref{fig:energy_drift}. Two facts are discernible from these plots: the energy drifts are approximately linear, and larger expensive potential regions lead to greater energy gain. The right hand side of Fig.~\ref{fig:energy_drift} shows the energy gains normalized by the (approximately spherical) surface interface area between potentials. Under this normalization, the curves collapse onto a single straight line, giving us the result that for two given potentials and a given blending regime the rate of energy increase per unit surface contact area is approximately constant. The gradient of this line can be thought of as the energy flux from the force mixing contact surface into the material.

\subsubsection{Flexibility of cheap potential}
\begin{figure*}
    \centering
    \includegraphics[width=0.9\linewidth]{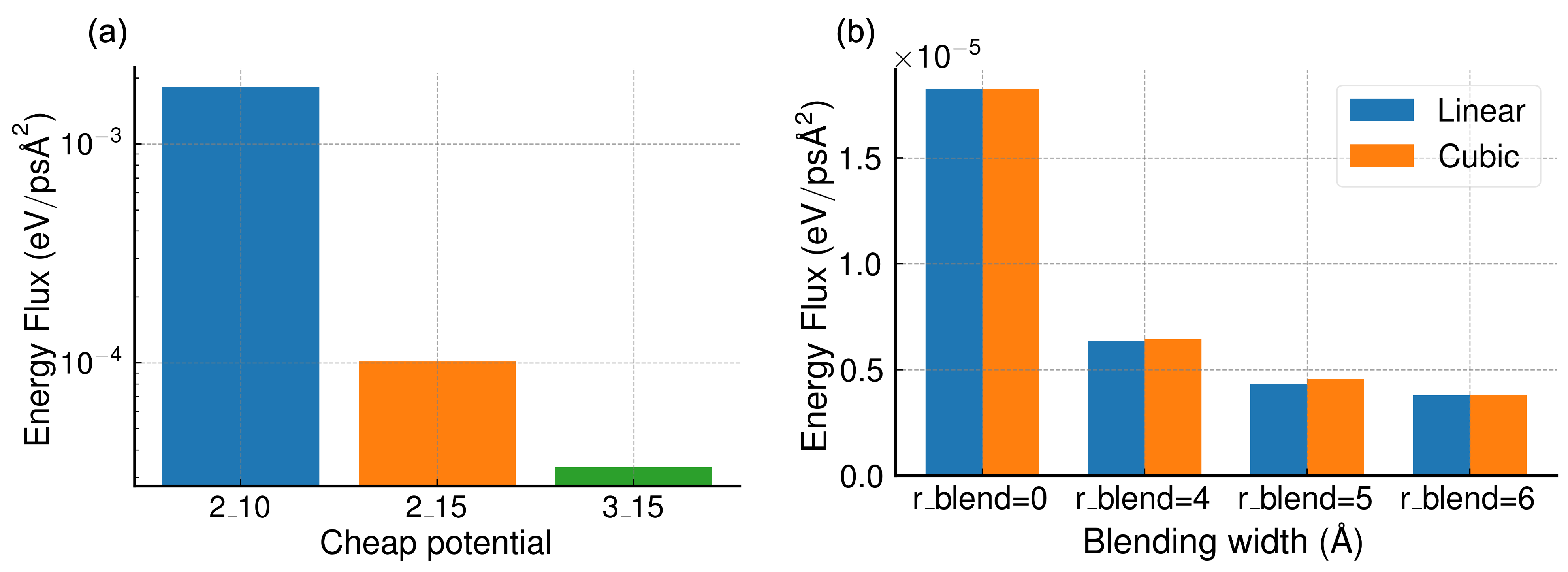}
    \caption{Measurements of (positive) energy flux from the ML/ML boundary for different cheap potentials and blending regimes. \TC{(a)}: Energy flux for abruptly mixed ML/ML simulations with different cheap potentials of increasing complexity. \TC{Blue, orange and green bars represent the measured flux for the 2\_10, 2\_15 and 3\_15 cheap potentials respectively.} More complex cheap potentials lead to orders of magnitude lower energy drift. (b): Energy flux for ML/ML simulations with the 2\_10 potential with different blending widths and blending functions. r\_blend=0 corresponds to abrupt mixing, whilst the other bars correspond to increasing amounts of blending. Increasing the blending width decreases the energy flux from the potential contact surface. \rebuttal{Very little difference in energy flux was observed between using a \TC{linear blending function (blue bars) or cubic blending function (orange bars).}}}
    \label{fig:energy_pot_and_blending}
\end{figure*}
Energy fluxes from ML/ML simulations of three different cheap potentials are shown on the left-hand side of Fig.~\ref{fig:energy_pot_and_blending}. Note that as the cheap potential complexity increases, the energy flux from the internal potential/potential contact surface decreases by orders of magnitude. The rationale for this is straightforward; more complex cheap potentials with greater numbers of basis functions have the flexibility to better approximate the local potential energy surface of the expensive potential.

\subsubsection{Width of blending region}
The right-hand side of Fig.~\ref{fig:energy_pot_and_blending} shows how adding a blending region in the ML-MIX simulation can decrease the energy flux. The \rebuttal{left-most} bars correspond to abrupt mixing, and it can be seen that as core region is swapped out for blending region the amount of energy drift in the simulation decreases. \rebuttal{The type of blending (linear or cubic) did not make any noticeable difference to measured energy flux.}

\printbibliography